\documentclass[twocolumn,floatfix,superscriptaddress,longbibliography,nofootinbib]{revtex4-1}
\RequirePackage[sort&compress]{natbib}

\usepackage{color}

\usepackage{amssymb,amsfonts,amsmath}
\usepackage{bm}
\usepackage[pdftex]{graphicx}
\usepackage{xcolor}
\usepackage{comment}

\usepackage{float}

\usepackage[colorlinks = true,
            linkcolor = blue,
            urlcolor  = blue,
            citecolor = blue,
            anchorcolor = blue]{hyperref}

\usepackage{dcolumn}
\usepackage{bm}

\usepackage[caption=false]{subfig}
\usepackage{xcolor}

\renewcommand{\(}{\left(}
\renewcommand{\)}{\right)}
\renewcommand{\[}{\left[}
\renewcommand{\]}{\right]}

\newcommand{\tr}[1]{\text{Tr}\(#1\)}
\renewcommand{\(}{\left(}
\renewcommand{\)}{\right)}
\newcommand{\smf}{S_{\text{MF}}}
\newcommand{\sint}{S_{\text{int}}}
\newcommand{\sintmf}{S_{\text{int},\text{MF}}}
\newcommand{\zint}{Z_{\text{int}}}
\newcommand{\snint}{S_{\text{nint}}}
\newcommand{\sagg}{S_{\text{agg}}}
\newcommand{\smean}{\bar{S}_{\text{nint}}}

\def\@cite#1#2{$^{\mbox{\scriptsize #1\if@tempswa , #2\fi}}$}

\newcommand{\heading}[1]{\noindent\vspace{0.1truecm}\textbf{#1.--}~}

\usepackage{xcolor}
\definecolor{RoyalBlue}{HTML}{4169e1}
\definecolor{ForestGreen}{HTML}{228b22}
\definecolor{DarkGreen}{HTML}{006400}

\usepackage{setspace}
\usepackage{comment}

\usepackage{todonotes}

\newcommand{\revv}[1]{#1}

\begin{document}


\title{Enhancing transport properties in interconnected systems without altering their structure}

\author{Arsham Ghavasieh}
\affiliation{Universit\'{a} degli Studi di Trento, Via Sommarive 5, 38123 Povo (TN), Italy} 
\affiliation{Fondazione Bruno Kessler, Via Sommarive 18, 38123 Povo (TN), Italy}

\author{Manlio De Domenico}
\email[Corresponding author:~]{mdedomenico@fbk.eu}%
\affiliation{Fondazione Bruno Kessler, Via Sommarive 18, 38123 Povo (TN), Italy}

\date{\today}

\begin{abstract}

 Units of complex systems -- such as neurons in the brain or individuals in societies -- must communicate efficiently to function properly: e.g.,  allowing electrochemical signals to travel quickly among functionally connected neuronal areas in the human brain, or allowing for fast navigation of humans and goods in complex transportation landscapes. The coexistence of different types of relationships among the units, entailing a multilayer represention in which types are considered as networks encoded by \emph{layer}s, plays an important role in the quality of information exchange among them. While altering the structure of such systems -- e.g., by physically adding (or removing) units, connections or layers -- might be costly, coupling the dynamics of subset(s) of layers in a way that reduces the number of redundant diffusion pathways across the multilayer system, can potentially accelerate the overall information flow. To this aim, we introduce a framework for \emph{functional reducibility} which allow us to enhance transport phenomena in multilayer systems by coupling layers together with respect to dynamics rather than structure. Mathematically, the optimal configuration is obtained by maximizing the deviation of system's entropy from the limit of free and non-interacting layers. Our results provide a transparent procedure to reduce diffusion time and optimize non-compact search processes in empirical multilayer systems, without the cost of altering the underlying structure.

\end{abstract}

\maketitle


\section{Introduction}

A wide variety of social, natural and artificial systems are inherently complex~\cite{boccaletti2006complex}. For instance, societies exhibit rich microscopic dynamics, at the level of single individuals, that might lead to emergent collective phenomena at larger scales~\cite{gabaix2003theory} -- e.g., financial collapses or revolutions --  which are usually difficult to predict~\cite{Bardoscia2017}. Natural systems, like ecological ones, are characterized by complex topologies that, in presence of interdependencies with other networks, 
exhibit a richer response to perturbations with respect to the case where they are considered in isolation~\cite{Pilosof2017}. Similar phenomena have been observed also for engineering systems, from transportation to communication networks, where their resilience to targeted attacks or random failures of their units is important for applications~\cite{Albert2000,de2014navigability,Gao2016}, such as robustness enhancement or recovery strategies~\cite{Majdandzic2013,majdandzic2016multiple}.

\begin{figure*}[!ht]
\includegraphics[width=\textwidth]{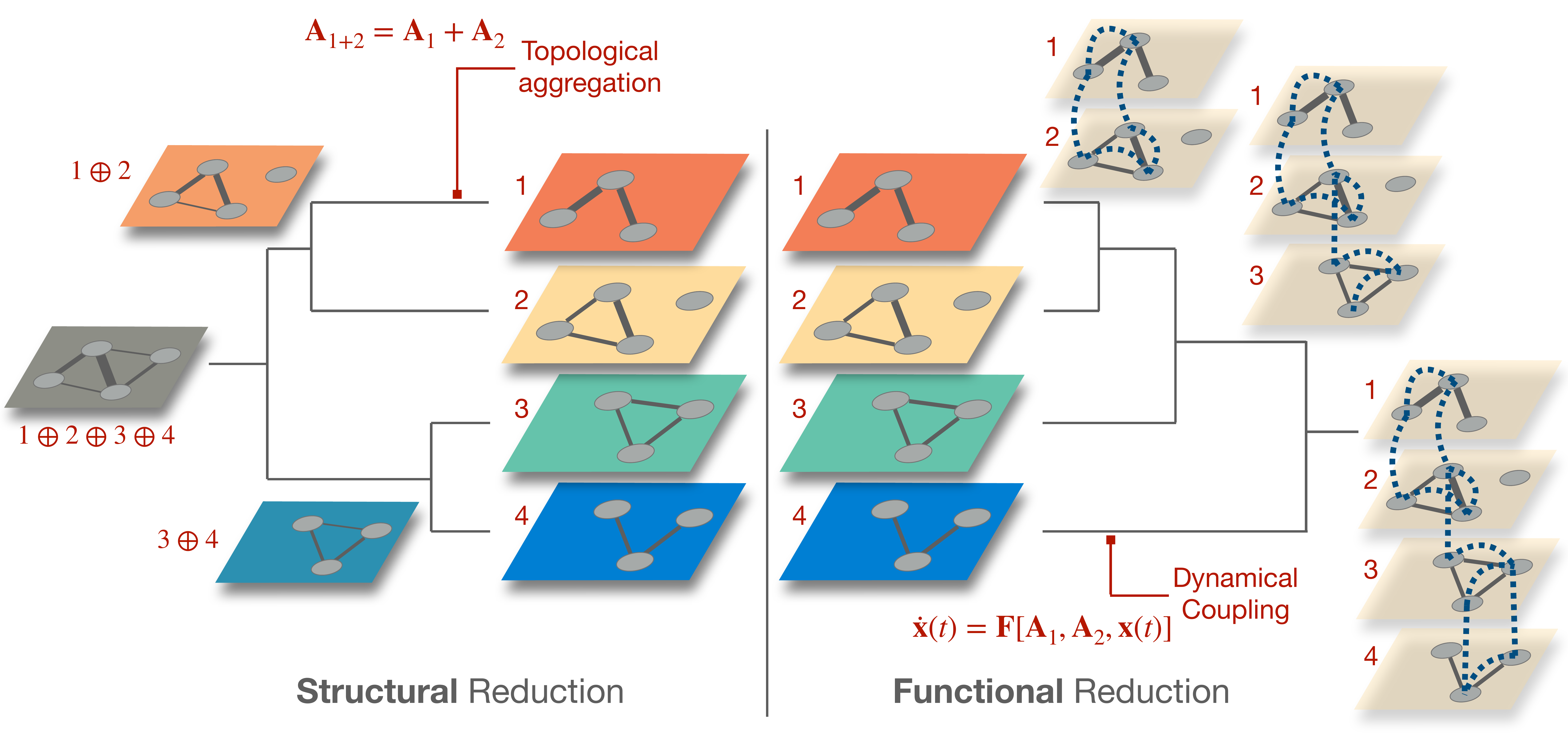}
\caption{\label{fig:functional_vs_structural}\textbf{Structural \emph{versus} functional reducibility of multilayer systems.} In structural reduction (left) one alters the structure by iteratively aggregating layers, e.g., by summing the corresponding adjacency matrices. However, this type of intervention usually comes with some costs (e.g., temporal, infrastructural, economic, etc.). Here, we propose functional reduction (right), an alternative approach where the structure of the system is not altered while coupling the dynamics on the top of layers. While any dynamics is allowed in principle, diffusive processes are particularly suitable ones because they allow for theoretical and computational treatment. In this work, random walk dynamics is considered and layers are functionally coupled if their network states are not distinguishable from the point of view of a walker (i.e., the same color is assigned to functionally coupled layers). From an application perspective, this approach corresponds to shared tickets for public multimodal transport infrastructure or shared promotions for users who are part of multiple social networks at the same time.}

\end{figure*}

In fact, complex systems are characterized by a wide variety of physical attributes and dynamics, making difficult to analyze them within a unified framework. What do electrochemical signals exchanged among neurons in the human brain, mobility of goods/people/vehicles between different areas of a urban ecosystem, spreading of an infectious pathogen through social contact patterns and financial transactions through the backbone of a stock market have in common? From a very general perspective, the corresponding systems consist of interconnected units exchanging \emph{information} similarly to how computers and servers exchange packets of bits through the Internet. In fact, it has been originally conjectured by Murray Gell-Mann that despite their differences, complex systems resemble one another in the way they handle information and, consequently, studying how information is exchanged and processed provides a promising starting point for exploring transport phenomena in order to understand how such systems operate.

A general way to model the propagation of information through complex structures is by means of diffusive processes, such as random walks~\cite{Masuda2017}, since they are versatile models that allows one to cope with the uncertainty in the structure -- e.g., temporary link failures or damages -- of complex systems, they are based on local knowledge of structure \cite{randomcom} -- which is usually less expensive than models relying only on the global knowledge of the topology, like shortest-path-based ones. Nevertheless, there is increasing evidence that units of real complex systems do not preferentially exchange information along shortest paths, too expensive in terms of routing. In fact, the human brain operates in an intermediate communication regime, neither based on shortest-path routing nor on a diffusion process~\cite{avena2019spectrum}, while the Internet relies on routing tables – more specifically, forwarding tables chosen by the routing algorithm – on each machine to find sub-optimal communication pathways through the routing hierarchy~\cite{rfc1009,rfc1812}. Even individual human mobility, usually assumed to be far from random~\cite{song2010limits} is affected by arbitrariness of individuals’ actions responsible for a stochastic component~\cite{smith2014refined} in trip patterns characterizing human flows~\cite{gallotti2019disentangling}, which is usually modeled by means of stochastic processes ranging from random walks to variants of brownian motion and Levy flights~\cite{barbosa2018human}. In this work, we present a way to alter the dynamics on top of multilayer networks, while keeping their structure (see Fig.~\ref{fig:functional_vs_structural}). Without changing the underlying structure, shortest paths across the whole multiplex system do not change accordingly. This unfortunate feature of shortest paths makes them useless to enhance transport properties under the hard constraint of keeping the structure unaltered.

Therefore, constituents of complex systems tend to or are designed for exchanging information in a very efficient way to function properly. An emblematic example is given by the human brain, where the collective dynamics of billions of neurons is responsible for coordinating the human body and cognitive activities, while keeping energy cost as low as possible~\cite{Bullmore2012}. However, a deeper understanding of how to act on the system to enhance or hinder transport phenomena -- such as navigability~\cite{boguna2009navigability} -- continues to elude us, because they depend on the interplay between structure and dynamics of a complex system, which is usually difficult to model and quantify~\cite{liu2011controllability}. Nevertheless, understanding such an interplay represents one of the most important challenges in complexity science, with promising advances. In fact, it has been recently shown that the relation between structure and dynamics of information flow can be unraveled by analyzing how perturbations propagate through the system~\cite{harush2017dynamic}. The complex interplay between structure and dynamics of information propagation has been also mapped to its latent geometry to better understand network-driven contagion phenomena~\cite{brockmann2013hidden} and, more recently, it has been shown how signal propagation is able to capture the role of network connectivity in propagating local information, thus linking the topology to the observed spatio-temporal spread of perturbative signals across it~\cite{hens2019spatiotemporal}.

However, the same task is even more challenging when the system of interest can be described in terms of a multilayer network~\cite{gao2012networks,de2013mathematical,gomez2013diffusion,Radicchi2013,Kivela2014,boccaletti2014structure,de2016physics}. For instance, multilayer models for urban and regional transportation systems  -- accounting for different means, from rails, to ships and flights, serving the same geographical areas simultaneously -- have highlighted that an efficient flow might be hindered by the lack of synchronization between different layers~\cite{gallotti2014anatomy}. 

Up to date, it is still unknown how to enhance flow distribution in multilayer systems, especially under constraints. For instance, one might want to add new fast connections (e.g., highways, tube, flights, etc.) between distant geographic areas to speed up the traffic flow. However, since each connection comes with an economic cost, it is very unlikely that most of them can be realized in practice. Similarly, one might avoid to cut existing connections, even if desirable in some cases, because they might have a high societal cost in terms of impact on the population. \revv{It has been recently shown that for multilayer networks with interlinks, changing the weights of interlinks can enhance the diffusion on top of the networks \cite{physrevlett110} and lead to the phenomenon of "super diffusion". However, in many complex systems, as the ones which can be modeled by coupled networks, interlinks either do not exist or it is not trivial to assign weights to them.} Therefore, the challenge is to enhance transport properties without altering the existing structure: this can be achieved, in principle, by functionally coupling together layers with similar flow patterns. \revv{ When each layer of the multilayer system is represented by a specific color, an emblematic way to understand functional coupling of two layers is to assign the same color to both of them and make them indistinguishable, as illustrated in Fig.~\ref{fig:functional_vs_structural}.}

It is worth remarking that this approach leaves the structures of layers intact, while only altering the dynamics on top of them, in a way that the dynamical process can not recognize functionally coupled layers as distinct ones. For instance, this type of solutions is adopted, to some extent, by airlines proposing shared flights to their customers.

Intuitively, one might think that reducing structural redundancy -- defined in terms of replicated connections across layers -- can provide a solution. Remarkably, the recent analysis of common estimators of structural redundancy, such as edge overlapping, revealed that the presence of redundant connectivity might boost the robustness of multiplex networks~\cite{radicchi2017redundant}. However, when more complex indicators of redundancy are needed, our understanding is dramatically more limited. On the one hand, recent studies focused on quantifying the distance between two complex networks in terms of their structural dissimilarities~\cite{schieber2017quantification} -- as a distance between probability distributions extracted from the networks -- or their spectral information content~\cite{de2016spectral} have been successfully used to compare classical, single-layer, networks. On the other hand, structural dissimilarities~\cite{carpi2018assessing} and spectral entropy divergence~\cite{de2015structural} between layers have been recently proposed as effective approaches for reducing the structure of such systems, based on different ordering strategies. However, methods relying only on structural features are not suitable for our purposes -- as they are costly in practice -- and they perform poorly, as we show later. The reason is that they are based on heuristics, lacking a fundamental understanding of the underlying physics.

We show that a fundamentally different perspective must be considered instead of structural redundancy: by using random walk dynamics as a proxy for information flow, transport phenomena are enhanced when subsets of layers are functionally grouped together in such a way that the corresponding diffusion pathways are maximally different. Therefore one must analyze the functional redundancy of a multiplex system, defined by abundant and redundant diffusion pathways that distribute the flow between different system's units.

In the following, for sake of clarity, to avoid any confusion with structural redundancy, we will refer to functional reduction and functionally reduced networks to indicate a loss of redundancy in diffusion pathways across layers. At variance with structural reducibility, functionally reducible multiplex systems consists of layers that can be functionally grouped into subsets to enhance transport, whereas irreducible systems do not allow for any functional aggregation. The difference between structural reducibility that alters the structure of multilayer networks, and the proposed functional reducibility that only alters the dynamics on top of multilayer networks, is illustrated in Fig.~\ref{fig:functional_vs_structural}.

In this work, we develop tools familiar to physicists -- such as partition function and mean-field calculations -- to investigate the interplay between structure and dynamics of multilayer systems from a spectral perspective -- an approach revealing successful for a similar purpose in the case of classical networks~\cite{arenas2006synchronization,gfeller2007spectral,krzakala2013spectral,castellano2017relating}. By exploiting the relation between the flow distribution in a multiplex network and the spectral diversity of its layers, our framework can provide a broad spectrum of applications -- from more optimal supply strategy that combine transportation of goods to more efficient navigability of urban areas by allowing for multimodal trips with the same ticket -- that can be achieved by devising tailored policies targeting the dynamics of systems, without the cost of changing their structure. 

\section{Statistical Physics of Random Walks in Complex Networks}

In this section, we introduce the state of multiplex networks, discuss the role of their structure in hindering the information flow, find an inequality relating partition functions of single layers to the partition function of the whole system and develop a mean-field approach to allow for analytical derivations of the next sections. Note that most mathematical details are provided in the Appendices, while we provide here the theoretical grounds required for defining and understanding the functional reducibility of multilayer systems.

\heading{Information flow in multiplex networks} Multiplex networks are a special class of multilayer systems providing a successful and widely used model for a broad variety of empirical systems~\cite{Kivela2014,boccaletti2014structure}. This class of networks is characterized by nodes with multiple types of relationships or interactions, simultaneously, which are encoded by layers. Here, physical nodes correspond to the physical units of the system, whereas state nodes correspond to the replicas of a physical node across layers.

Information flow in complex networks has been successfully modeled by diffusive processes such as random walks~\cite{Masuda2017}. However, while random walk dynamics has been previously introduced for interconnected multilayer systems~\cite{de2013mathematical,de2014navigability}, the case of multiplex systems has been recently studied only by using the full aggregation of the system into a single-layer network~\cite{battiston2016efficient}. Clearly, the dynamics of a random walker on the aggregated representation of a multiplex system can not coincide with the multiplex dynamics. 

To describe the random walk dynamics on top of multiplex networks, we first find the corresponding transition matrix, that is given by $\mathbf{T}=\langle\mathbf{T}^{(\ell)}\rangle $ (See Appendix~\ref{app:randommultiplex}), and encodes the probability of jumps from each node to any other. Similarly, the normalized Laplacian matrix corresponding to the multiplex random walk dynamics can be obtained as $\boldsymbol{\mathcal{L}}=\mathbf{I}-\mathbf{T}=\langle \boldsymbol{\mathcal{L}}^{(\ell)} \rangle$, being $\mathbf{I}$ the identity matrix. This means that the normalized Laplacian matrix encoding random walks on a multiplex network is the weighted average of Laplacian matrices of layers. Importantly, each layer's weight corresponds to its contribution to the overall flow of information. As these weights are often hard to assess or not known at all, when this is the case we assume that there is no preference on layers, and from now on we consider uniform distribution of weights, $1/L$, for a multiplex network of $L$ layers. Nevertheless, it is worth remarking that the proposed method and the theoretical framework are valid for any general distributions of weights.

The propagator (see Appendix~\ref{app:randommultiplex}) governs the temporal evolution of the random walk and, for this reason, it encodes information about the interplay between the structure of the system and the random search dynamics on the top of it. In fact, it has been recently used to define the state of the system~\cite{de2016spectral}, similar in spirit to the approach widely used in quantum statistical mechanics to define the mixed state of entangled quantum units. In this framework, the network state is defined by $\boldsymbol{\rho}(\tau) = \frac{e^{-\tau \boldsymbol{\mathcal{L}}}}{Z}$ , where $Z(\tau)=\tr {e^{-\tau \boldsymbol{\mathcal{L}}}}$ is a normalization factor, playing the same role of the partition function of the system. This state, which is formally identical to a density matrix~\cite{RevModPhys.29.74}, is then used to calculate the Von Neumann entropy of a complex network as $ S(\tau)=-\tr{\boldsymbol{\rho}(\tau)  \log_{2} \boldsymbol{\rho}(\tau) }$. 
See Appendix~\ref{app:notation} for further details.

\heading{Physical meaning of the partition function} Here, we provide first a deeper understanding of the meaning of the partition function. In fact, $Z(\tau)=N\mathcal{R}(\tau)$, being $\mathcal{R}(\tau)=N^{-1}\sum\limits_{i=1}^{N}e^{-\tau \lambda_{i}}$ the average return probability of random walker and $\lambda_{i}$ is the $i$--th eigenvalue of $\boldsymbol{\mathcal{L}}$   (see Appendix~\ref{app:dynamicaltrapping}). As mentioned before, random walk can be used as a proxy of information transport in complex networks. Intuitively, high average return probability is associated with the tendency of structure to trap the information flow in its initial place. So, this correspondence between partition function of the system and average return probability, unravels how partition function relates to transport phenomena in complex networks. Furthermore, one can define density matrices in terms of any valid Laplacian encoding other types of dynamics (e.g., continuous diffusion). In this case, while the average return probability can not even be defined, the partition function still exists and corresponds to the amount of information that is trapped in its initial place and flow more difficultly towards peripheral parts of the network. 

As mentioned above, high average return probability indicates that the random walker can not explore efficiently the rest of the network. Consequently, the presence of structural symmetries and abundance of redundant diffusion pathways for information to propagate between system's units are expected. For instance, in an ordered and symmetric structure like a regular lattice, where each node interacts only with its first neighbors and no long-range interactions are possible, the information exchange between distant nodes is slow and inefficient. Conversely, low average return probability indicates that the random walker can navigate the network faster. For instance, breaking down structural order by adding long range interactions between distant nodes, generating topological shortcuts and other defects like in small-world~\cite{Watts1998} and scale-free~\cite{Barabasi1999} networks, information flows faster and transport properties are enhanced, accordingly. $Z(\tau)$, as a measure of transport that we name average dynamical trapping in the following, is able to encode these phenomena. The same arguments apply to multiplex networks, with the difference that in this case we are wondering if the redundancy of diffusion pathways across layers can be reduced by superimposing subsets of layers, in order to reduce the average dynamical trapping. 

\begin{figure*}[!ht]
\centering
\includegraphics[width=\textwidth]{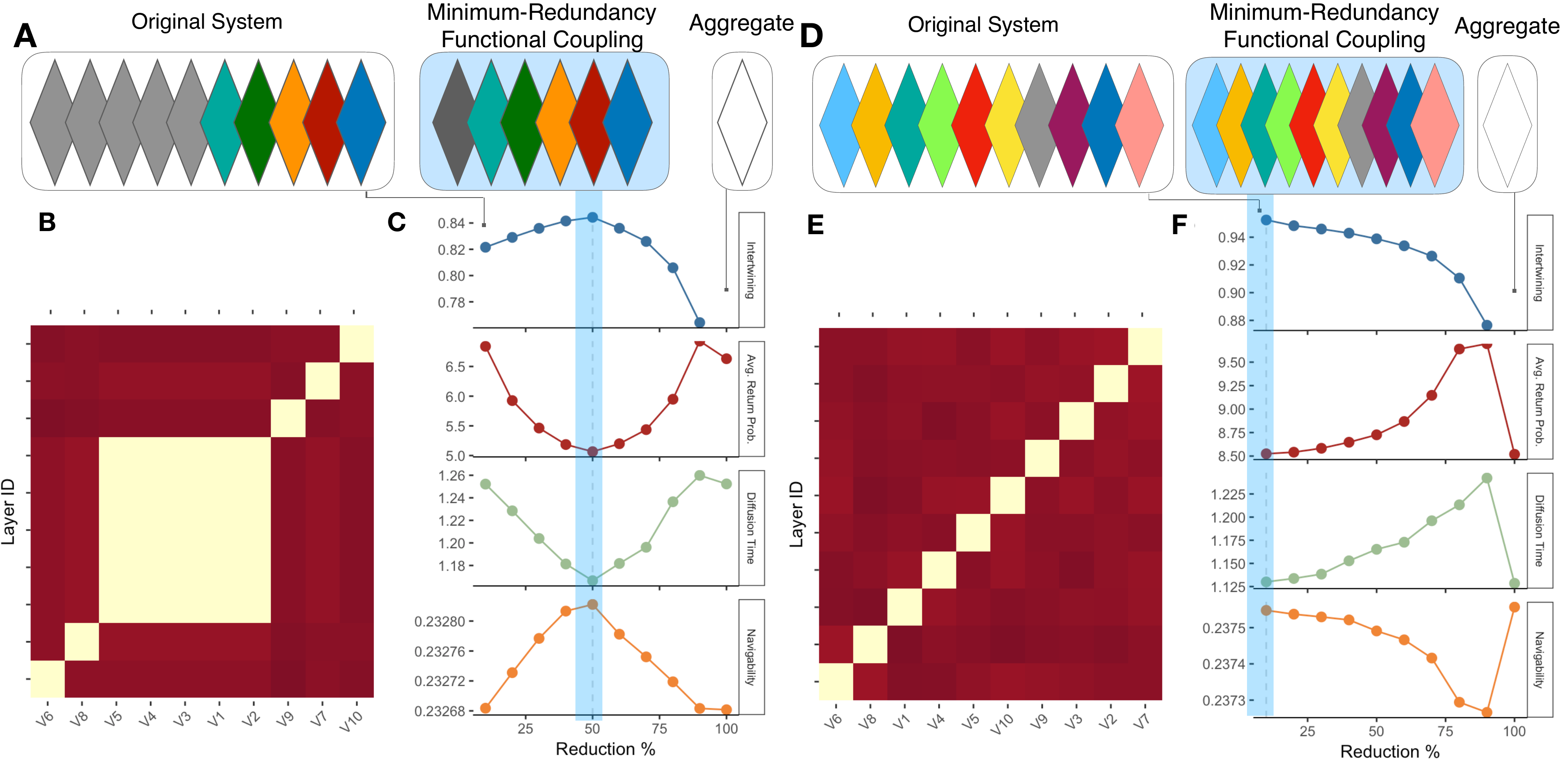}
\caption{\label{fig:synthetic}\textbf{Functional reducibility in synthetic systems} (A) A multiplex network with 100 nodes, consisting of 5 identical layers and 5 distinct random networks with wiring probability 0.2 is illustrated in its simulated form (Original System), its form after functionally coupling its layers according to the framework presented in this study (Minimum-Redundancy Functional Coupling) and the functional form corresponding to the coupling of all layers simultaneously (Aggregate). (B) Heatmap encoding pairwise dissimilarity (the darker the color, the larger the distance) between layers shown in panel (A), according to collective cosine distance (see Appendix \ref{app:CCD}). For clarity, cells are re-arranged in such a way that closer layers have smaller distance. (C) From top to bottom: intertwining compared to transport properties such as average return probability \revv{(rescaled by $N$)}, diffusion time and navigability, respectively, that can be enhanced on this multiplex network by using functional reducibility. Note that average return probability $\mathcal{R}(m)$ at reduction step $m$, integrated over $\tau$, has been then transformed by $-\log(1-\mathcal{R}(m)/\max\limits_{m}\{\mathcal{R}(m))\}$ for clarity. (D) A multiplex network with 100 nodes where layers are uncorrelated random networks with wiring probability 0.2. (F) As in (B), but for the system considered in panel (D). \revv{(F)} Transport properties are not enhanced by functional coupling because spectral diversity is maximized when the original system is not functionally aggregated. See Fig.~\ref{fig:app:synthetic} for results on additional synthetic models.}
\end{figure*}

\heading{Fundamental inequality for multiplex systems} Multiplexity of interactions among units generate non-trivial dynamical correlations between layers that have no counterpart when layers are considered in isolation. This important difference can be characterized in terms of differences in information content, quantified by average entropy distance of the multiplex and its layers (see Appendix~ \ref{app:partition}), that we call \emph{intertwining}. In the next sections, we discuss the central role of intertwining in functional reducibility. 

Directly from intertwining, a fundamental inequality between the partition function of a multiplex system as whole and the partition functions of its layers can be derived. This inequality is important, as it relates the transport phenomena of multiplex system and layers, through average dynamical trapping (see Appendix~\ref{app:partition} for details):
\begin{eqnarray}
\label{eq:fundrel}Z(\tau) \leq \prod\limits_{\ell=1}^{L} Z^{(\ell)}(\tau)^{1/L},
\end{eqnarray}
where equality holds if and only if all the layers are the same, i.e. the random exploration of the network does not depend on any specific layer of the system. This equality defines the non-interacting scenario where layer-layer correlations do not alter the underlying dynamics: the entropy $S^{(\ell)}(\tau)$ of each layer is calculated separately and the overall entropy is given by their average $\snint(\tau)=\langle S^{(\ell)}(\tau)\rangle$. Conversely, any topological alteration of the non-interacting scenario introduces a dynamical correlation between layers, requiring the exploration of layers to gather more information about the system: in this case the network consists of interacting layers, where the diffusion dynamics on the whole multiplex network is considered to measure the entropy $\sint(\tau)$.

\heading{Mean-field entropy} An analytical treatment of Von Neumann entropy is, in general, difficult and we have developed a spectral mean-field theory to cope with this complexity (Appendices~\ref{app:mf}--\ref{app:mf1st}):
\begin{eqnarray}
\label{eq:entrmf}
\smf(\tau) = \frac{1}{\log 2}\(\tau\frac{Z(\tau)-1}{Z(\tau)}+\log Z(\tau)\)
\end{eqnarray}
In the next section, we use mean-field entropy to find a simple representation of intertwining, in terms of $\sint$ and $\snint$.
\section{Quantifying layer-layer interactions: the intertwining}

Using fundamental inequality~(\ref{eq:fundrel}) and mean-field entropy(\ref{eq:entrmf}), it can be shown that $\sint(\tau)\leq \snint(\tau)$ (Appendix~\ref{app:intertwining}): i.e., layer-layer interactions lower system's  entropy and, consequently, its partition function $Z(\tau)$. We exploit the average information divergence of layers from the whole (see Eq.~\ref{eq:avgdiv}) rescaled by its upper bound, to provide a normalized measure that quantifies the importance of describing the system in terms of coupled networks rather than networks in isolation. For values of $\tau$ sufficiently large, and in absence of isolated state nodes, it can be shown that the relative intertwining reduces to the deviation between interacting and non-interacting entropies (see Appendix~\ref{app:intertwining_dimred} for details):
\begin{eqnarray}
\mathcal{I}^{*}(\tau) =1 -\frac{\sint(\tau)}{\snint(\tau)},
\end{eqnarray}
which is bounded between 0 (i.e., the system is reducible to a single network) and 1 (i.e., the system is irreducible).

Before presenting how to use the relative intertwining for the analysis of multiplex networks, we derive analytically its direct dependence on the partition function and the eigenvalues of the normalized Laplacian matrix, especially on $\lambda_{2}$ -- the second smallest one -- which governs many transport phenomena, as shown in the following. From the master equation, the Laplacian matrix of the multiplex network is given by $\boldsymbol{\mathcal{L}}=\langle \boldsymbol{\mathcal{L}}^{(\ell)}\rangle$: here, we write the Laplacian matrix $\boldsymbol{\mathcal{L}}^{(\ell)}$ of each layer as the perturbation $\boldsymbol{\mathcal{L}}^{(\ell)}=\boldsymbol{\mathcal{L}}+\Delta \boldsymbol{\mathcal{L}}^{(\ell)}$, reflecting in the corresponding eigenvalues as $\lambda_{i}^{(\ell)}=\lambda_{i} + \Delta\lambda_{i}^{(\ell)}$ $(i = 0,1,...N)$. 

It can be shown that $\langle\Delta\lambda^{(\ell)}\rangle=0$ (see Appendix~\ref{app:intertwining_dimred}) and that $\langle (\Delta\lambda^{(\ell)})^{2} \rangle \geq 0 $, the latter quantifying the influence of the perturbation to each layer. The variance is zero only in the case of identical layers. 

A quantity of interest is the variance averaged across all layers, i.e. $\overline{\left\langle\(\Delta\lambda^{(\ell)}\)^{2} \right\rangle}$, which measures the overall influence of the perturbation: it is expected to increase for increasing deviation of layers from the average, i.e. for increasing diversity of the layers. In fact, we show that the relative intertwining is directly proportional to this variance (Appendix~\ref{app:intertwining_dimred}):
\begin{eqnarray}
\label{eq:ivsvariance}
\mathcal{I}^\star(\tau) &\approx& \frac{\tau^{2}}{2} \overline{\left\langle\(\Delta\lambda^{(\ell)}\)^{2} \right\rangle},
\end{eqnarray}
which increases as the diversity of diffusion pathways of layers increases.

Similarly, $Z^{(\ell)}(\tau)=\zint(\tau)+\Delta Z^{(\ell)}(\tau)$ holds for partition functions and we show that (see Appendix~\ref{app:intertwining_dimred} for details)
\begin{eqnarray}
\label{eq:ivsz}
\mathcal{I}^{\star}(\tau) &\approx& \frac{\overline{\Delta Z^{(\ell)}}(\tau)}{ \zint(\tau) - 1 };\ \overline{\Delta Z^{(\ell)}}(\tau)=\frac{1}{L}\sum\limits_{\ell=1}^{L} \Delta Z^{(\ell)}(\tau).
\end{eqnarray}
Equations~(\ref{eq:ivsvariance}) and (\ref{eq:ivsz}) provide the first fundamental result of this work: they show that by minimizing the partition function of the system one maximizes the relative intertwining while favoring the maximum functional diversity of layers. The second fundamental result of this work is the direct relationship between the relative intertwining and the second smallest eigenvalue $\lambda_2$ of the normalized Laplacian matrix (see Appendix~\ref{app:intertwining_transport} for details):
\begin{eqnarray}
\mathcal{I}^{\star}(\tau)&\approx& \tau(\lambda_{2} - \overline{\lambda_{2}^{(\ell)}} ).
\end{eqnarray} 
This relation is important because it highlights how maximizing intertwining corresponds to maximize $\lambda_2$, which in turn is equal to the inverse of diffusion time and it plays a crucial role for navigability as two important transport properties of complex networks (see Appendix \ref{app:transport}).

Thus, by maximizing the relative intertwining, one actually enhances the most important transport phenomena on multiplex systems and, as a byproduct, reduces their dimensionality. To use our framework for practical applications, we need a strategy for identifying similar layers in a multiplex network, aggregating them accordingly, and a stopping criterion. We use a novel approach, namely collective cosine distance, which is computationally faster and more reliable than existing methods when dealing with isolated nodes and multiple connected components (see Appendix \ref{app:CCD}) to cluster layers. Note that we perform an exhaustive pairwise search among possible functional groups, \emph{de facto} getting rid of heuristics such as hierarchical clustering~\cite{de2015structural,de2016spectral}.

\section{Functional reducibility of multiplex systems}

For each value of the time $\tau$, we calculate all pairwise distances between layers to to find the most similar ones: at each step $m$ of this sequence, the corresponding value $\mathcal{I}^{\star}(\tau;m)$ of the relative intertwining is calculated. Finally, the values of relative intertwining are averaged over $\tau$ as $\mathcal{I}^{\star}(m)=\langle\mathcal{I}^{\star}(\tau;m)\rangle$ and used for comparison against navigability, diffusion time at each aggregation step. We repeat this procedure until the original system is aggregated into a single network, representing the superposition of the whole system from the dynamical perspective. The average relative intertwining is expected to reach a maximum value if any desirable superposition of subsets of layers exists (see Appendix~\ref{app:intertwining_dimred} analytical details): in this case, the resulting system is characterized by enhanced transport properties. At this optimal grouping, in the sense that it is the most desirable configuration among the possible ones, the remaining number of layers can be indicated by $L_{\text{opt}}$ and one can define the \emph{reducibility} of a multilayer network as $\chi=(L-L_{\text{opt}})/(L-1)$ to characterize this system's feature with one scalar. Note that $\chi$ quantifies the fraction of layers grouped together: one can use either structural reducibility~\cite{de2015structural} -- maximizing the topological distinguishability from the aggregate --  or functional reducibility -- maximizing differences in diffusion pathways across layers -- defined in this work. However, when transport phenomena are a concern, functional reducibility must be preferred.

\heading{Synthetic systems} We have performed several numerical experiments to validate our theory on toy models with functionally reducible or irreducible systems. We show one representative case for each case in Fig.~\ref{fig:synthetic}. For the reducible case, see Fig.~\ref{fig:synthetic}A--C, our framework is able to detect functional redundancy and reduce it as expected, while lowering diffusion time, average return probability and dynamical trapping, and increasing navigability. For the irreducible case, see Fig.~\ref{fig:synthetic}D--F, a multiplex system consisting of noisy and independent layers is not functionally reducible, as expected. For further synthetic systems, see Appendix~\ref{app:syntheticapp}.

Remarkably, all transport properties considered here are enhanced by functionally coupling layers according to our theoretical framework.

\begin{figure*}[!t]
\centering
\includegraphics[width=\textwidth]{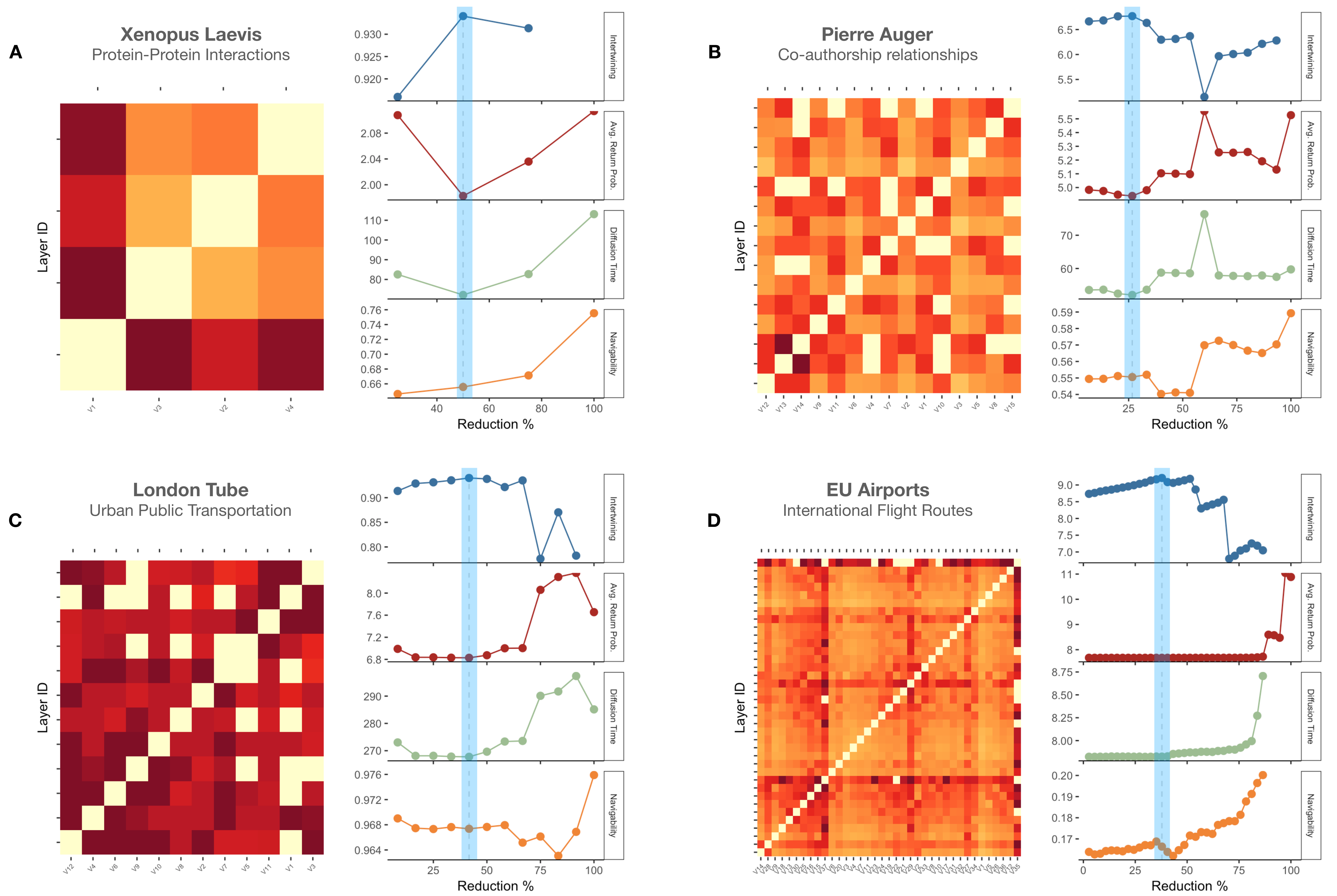}
\caption{\label{fig:empirical}\textbf{Functional reducibility of empirical multilayer systems} The same analysis shown in Fig.~\ref{fig:synthetic} is performed on different multilayer systems, namely the: (A) protein-protein interactions of \emph{Xenopus Laevis} (the African clawed frog)~\cite{de2015structural}; (B) co-authorship network of the Pierre Auger experiment~\cite{de2015identifying}; (C) public transportation of London~\cite{de2014navigability}; (D) European airport network~\cite{cardillo2013emergence}. See the text for further details about the data set. Note that entries $d_{ij}$ of distance matrices have been transformed by $-\log(d_{ij})$ to enhance the visualization of the corresponding heatmaps.\revv{(The average return probability is rescaled by $N$)}} 
\end{figure*}

In the following, we apply our framework to analyze the functional reducibility of empirical multilayer systems, including biological, social and transportation ones. 

\heading{Biological system} As a representative biological system, we consider the largest connected component of molecular interactions of 263 proteins characterizing the proteome of \emph{Xenopus Laevis} (the African clawed frog)~\cite{de2015structural}, where layers encode different genetic interactions (Association, Direct interaction, Physical association, Colocalization and Suppressive genetic interaction defined by inequality)~\cite{stark2006biogrid}. Note that the last layer consists only of one link and we discard it from the analysis. Results are shown in Fig.~\ref{fig:empirical}A and summarized in Tab.~\ref{tab:Xenopus}.

\begin{table}[!ht]
\begin{tabular}{c|c}
\hline
\textbf{functional group} & \textbf{layer ID (layer name)}                                                  \\ \hline
1                         & \begin{tabular}[c]{@{}c@{}}1 (Association) \\ 4 (Colocalization)\end{tabular} \\ \hline
2                         & 2 (Direct interaction)                                                          \\ \hline
3                         & 3 (Physical association)                                                        \\ \hline
\end{tabular}
\caption{Functional coupling of layers in the multiplex protein-protein interaction network of \emph{Xenopus Laevis}. Reducibility: $\chi_{func}=33.3\%. $}
\label{tab:Xenopus}
\end{table}

Our analysis suggests that, from a functional perspective, protein-protein interactions characterized by Association and Colocalization in this organism can be coupled together. This result is in striking disagreement with the result obtained from structural reducibility, where the other two layers were aggregated because indistinguishable from an information-theoretic perspective~\cite{de2015structural}.

\heading{Social system} The social system considered is the largest connected component of the collaboration network of the 475 scientists affiliated to the Pierre Auger experiment~\cite{de2015identifying}, the largest
observatory of ultra-high-energy cosmic rays. Here, collaborators work together in different research topics
on specific tasks defining the 16 layers (e.g., Source detection, Mass composition, Experimental enhancements, Shower reconstruction, etc) of the multilayer co-authorship system. Note that one layer was so sparse with respect to the other layers that we have discarded it from the analysis. Results are shown in Fig.~\ref{fig:empirical}B and summarized in Tab.~\ref{tab:Auger}.

\begin{table}[h!]
\begin{tabular}{c|c}
\hline
\textbf{functional group} & \textbf{layer ID (layer name)}                                                             \\ \hline
1                        & \begin{tabular}[c]{@{}c@{}}5 (Point-source),\\  2 (Detector), \\ 9 (Spectrum)\end{tabular} \\ \hline
2                        & \begin{tabular}[c]{@{}c@{}}6 (Mass-composition), \\ 12 (SD-reconstruction)\end{tabular} \\ \hline
3                        & 1 (Neutrinos)   \\ \hline
4                        & 3 (Enhancements)\\ \hline
5                        & 4 (Anisotropy)  \\ \hline
6                        & 7 (Horizontal)  \\ \hline
7                        & 8 (Hybrid-reconstruction) \\ \hline
8                        & 10 (Photons) \\ \hline
9                        & 11 (Atmospheric)\\ \hline
10                        & 13 (Hadronic-interactions) \\ \hline
11                       & 14 (Exotics) \\ \hline
12                       & 15 (Magnetic)   \\ \hline
\end{tabular}
\caption{Similar to Tab.~\ref{tab:Xenopus}, for the Pierre Auger collaboration network. Reducibility: $\chi_{func}=21.4\%$.}
\label{tab:Auger}
\end{table}

In this case, our analysis suggests the existence of two functional groups of layers. The two groups do not reflect similarities observed in the multilayer community organization of scientists~\cite{de2015identifying}, highlighting how the suggested functional coupling does not reflect trivial or complex structural features.

\heading{Urban and large-scale transportation systems} We further consider two transportation systems at different scales: the urban public transportation system of London consisting of 11 tube lines, DLR and Overground~\cite{de2014navigability} and 369 stations, and the European airport network~\cite{cardillo2013emergence}, consisting of 417 airports connected by flight routes served by 37 airlines defining layers. Results are shown in Fig.~\ref{fig:empirical}C--D and summarized in Tab.~\ref{tab:EUairline}--\ref{tab:London}, respectively.

Remarkably, in the case of London transportation, the result differs from the one obtained from structural reducibility, which instead predicts the aggregation of only one layer ($\chi_{struct}=8\%$~\cite{de2015structural}). This effect is due to the fact that distinguishability with respect to the aggregate network is not sufficient, alone, to capture the complex effects due to layer-layer coupling which, instead, are perfectly accounted for by the system's intertwining. In fact, the functional reducibility $\chi$ obtained in this case is 33.3\%, with the advantage of lowering both the diffusion time and dynamical trapping. Interestingly, DLR and Overground, which are very different from other tube lines, are not functionally coupled with other layers, confirming that, from a functional perspective, they are not redundant for flowing passengers.

\begin{table}[h!]
\begin{tabular}{c|c}
\hline
\textbf{functional group} & \textbf{layer ID (layer name)}                                                   \\ \hline
1                         & \begin{tabular}[c]{@{}c@{}}2  (jubilee)\\ 11 (circle)\end{tabular}               \\ \hline
2                         & \begin{tabular}[c]{@{}c@{}}4  (hammersmith and city)\\ 10 (central)\end{tabular} \\ \hline
3                         & \begin{tabular}[c]{@{}c@{}}6  (bakerloo)\\ 12 (victoria)\end{tabular}            \\ \hline
4                         & \begin{tabular}[c]{@{}c@{}}7  (piccadilly)\\ 8  (northern)\end{tabular}          \\ \hline
5                         & 1  (dlr)                                                                         \\ \hline
6                         & 3  (district)                                                                    \\ \hline
7                         & 5  (overground)                                                                  \\ \hline
8                         & 9  (metropolitan)                                                                \\ \hline
9                         & 13 (waterloo and city)                                                           \\ \hline
\end{tabular}
\caption{Similar to Tab.~\ref{tab:Xenopus}, for London Tube. Reducibility: $\chi_{func}=33.3\%$.}
\label{tab:London}
\end{table}

\begin{table}[!h]
\begin{tabular}{c|c}
\hline
\textbf{functional group} & \textbf{layer ID (layer name)} \\ \hline
1                         & \begin{tabular}[c]{@{}c@{}}2 (Ryanair)\\ 27(Transavia Holland)\\ 15 (Flybe)\end{tabular}    \\ \hline
2                         & \begin{tabular}[c]{@{}c@{}}7 (Air France)\\ 
11 (Swiss International Air Lines)\end{tabular}  \\ \hline
3                         & \begin{tabular}[c]{@{}c@{}}8 (Scandinavian Airlines)\\ 20 (LOT Polish Airlines)\end{tabular} \\ \hline
4                         & \begin{tabular}[c]{@{}c@{}}10 (Alitalia)\\ 23 (Air Lingus)\end{tabular}                      \\ \hline
5                         & \begin{tabular}[c]{@{}c@{}}12 (Iberia)\\ 17 (TAP Portugal)\end{tabular}                      \\ \hline
6                         & \begin{tabular}[c]{@{}c@{}}13 (Norwegian Air Shuttle)\\ 28 (Niki)\end{tabular}               \\ \hline
7                         & \begin{tabular}[c]{@{}c@{}}16 (Wizz Air)\\ 25 (Panagra Airways)\end{tabular}                 \\ \hline
8                         & \begin{tabular}[c]{@{}c@{}}18 (Brussels Airlines)\\ 32 (European Air Transport)\end{tabular} \\ \hline
9                         & \begin{tabular}[c]{@{}c@{}}19 (Finnair)\\ 33 (Malev Hungarian Airlines)\end{tabular}         \\ \hline
10                        & \begin{tabular}[c]{@{}c@{}}21 (Vueling Airlines)\\ 30 (Aegean Airlines)\end{tabular}         \\ \hline
11                        & \begin{tabular}[c]{@{}c@{}}24 (Germanwings)\\ 36 (TNT Airways)\end{tabular}                  \\ \hline
12                        & \begin{tabular}[c]{@{}c@{}}31 (Czech Airlines)\\ 34 (Air Baltic)\end{tabular}                \\ \hline
13                        & 1 (Lufthansa)                                                                                \\ \hline
14                        & 3 (Easyjet)                                                                                  \\ \hline
15                        & 4 (British Airways)                                                                          \\ \hline
16                        & 5 (Turkish Airlines)                                                                         \\ \hline
17                        & 6 (Air Berlin)                                                                               \\ \hline
18                        & 9 (KLM)                                                                                      \\ \hline
19                        & 14 (Austrian Airlines)                                                                       \\ \hline
20                        & 22 (Air Nostrum)                                                                             \\ \hline
21                        & 26 (Netjets)                                                                                 \\ \hline
22                        & 29 (SunExpress)                                                                              \\ \hline
23                        & 35 (Wideroe)                                                                                 \\ \hline
24                        & 37 (Olympic Air)                                                                             \\ \hline
\end{tabular}
\caption{Similar to Tab.~\ref{tab:Xenopus}, for the EU airport network. Reducibility: $\chi_{func}=35.1\%$.}
\label{tab:EUairline}
\end{table}

In the case of the EU airport network we observe a functional reducibility of about 35\%. It is worth mentioning that the analysis of structural reducibility of a larger data set -- encoding flight routes in Europe from 175 airlines serving 1064 airports -- reported a reduction of 6\%~\cite{de2015structural}: we will discuss better this result in the next section.

It is worth discussing about results obtained for the navigability of all empirical systems considered here. In fact, this transport descriptor is not enhanced, at variance with the other descriptors, when intertwining is maximized. However, this is not a limitation of the proposed framework because the equation relating intertwining to navigability has been obtained under an approximation that, evidently, is not satisfied when many isolated state nodes are present. In fact, even the coverage, and consequently the navigability, has been originally introduced to deal with interconnected multilayer systems where state nodes exist in all layers. In this work, we have introduced its generalization to the case of non-interconnected multiplex networks: when state nodes are defined in all layers, as in our synthetic benchmarks, both the coverage and the navigability behave as expected and, in fact, are correctly enhanced. When a large fraction of state nodes are present only in a few layers, as in most empirical systems, random walkers are more unlikely to reach those nodes, dramatically reducing the overall probability of hitting the corresponding physical nodes. Since the study of a new measure of coverage and navigability is beyond the scope of this work, we leave it for future studies devoted to better define such measures to deal with more exotic topological scenarios.

\section{Conclusions and outlook}

This study provides a transparent framework to better understand the interplay between complex topologies and non-compact search dynamics~\cite{chupeau2015cover}, allowing one to exploit the multifaceted interactions among constituents of complex systems -- usually encoded by multilayer networks -- to enhance transport phenomena without altering the underlying structure. 

First, we defined the average dynamical trapping of a complex network to quantify the efficiency in information exchange between its constituents: the flow distribution is more efficient when this trapping is minimized. Therefore, we demonstrated that this concept is intimately related to the spectral partition function of a complex network, allowing us to map the analysis of transport phenomena into the analysis of system's spectral entropy.

By applying the same principle to multilayer networks, we have identified a strategy for coupling subsets of layers -- according to a functional criterion -- in order to minimize the overall partition function. In fact, layers are grouped together by considering their superposition with respect to random walk dynamics. The choice of layers to superpose is driven by the deviation of system's entropy from the limit of non-interacting layers. This approach, justified from an information-theoretic perspective, has an elegant physical meaning: redundant diffusion pathways across layers are reduced until the multilayer system consists of layers with very different transport properties. This results is confirmed theoretically, showing that the average variance of the corresponding eigenvalue spectra must be maximum.

Results from synthetic benchmarks were in agreement with our expectations. Similarly, results from empirical multilayer systems confirmed the possibility to functionally couple group of layers to enhance information flow in the corresponding systems.

For instance, in the case of the collaboration network of Pierre Auger scientists, we provided evidence that some tasks should be coupled together. A mailing list -- or any other shared communication mean -- involving all scientists collaborating on the corresponding tasks would facilitate exchange of relevant information among peers. In fact, at the time described by the data set (i.e., between 2010 and 2012), an independent mailing list was assigned to each task and scientists had to subscribe to multiple ones to be updated with relevant information. However, this approach was not really efficient: it was not unusual to receive multiple times the same information.

The case of the European airport network was interesting as well. From a policy perspective, the functional coupling of routes corresponding to different airlines suggests that passenger flows would benefit from shared tickets. Among the identified functional groups, many are very reasonable when the location of the corresponding headquarters is considered: e.g., Air France and Swiss International Air Lines, Iberia and TAP Portugal, Ryanair and Flybe. Remarkably, while airline companies tend to avoid overlapping routes, as demonstrated by a structural reducibility of 6\%~\cite{de2015structural}, they are correlated (functional reducibility of 35\%) in a such a way that, from a functional point of view, they can still improve their transportation service. 

In the case of London public transport, we have found that a few configurations close to the functionally optimal one are also plausible in terms of enhanced transport properties and intertwining. This result was not surprising: the layers of this multiplex system have tree-like structures which tend to avoid topological overlap towards peripheral areas, thus different functional configurations might enhance transport in a similar way.

On the one hand, our findings open the doors to a broad spectrum of applications where superimposing layers with tailored policies induces relevant and controllable changes in transport phenomena, reducing diffusion time and enhancing navigability in multilayer networks. Nevertheless, the proposed approach can be easily adapted to scenarios where one physically acts on system's topology, not necessarily multilayer, for instance by adding (or removing) either its constituents or its connections. On the other hand, the theoretical framework developed in this work can be used further explore the bridge between statistical physics and network information theory.

\appendix

\section{Basic notation}\label{app:notation}

Given a complex network of $N$ nodes, undirected and weighted, the corresponding normalized Laplacian matrix governing the random walk dynamics, is defined by 
\begin{eqnarray}
\boldsymbol{\mathcal{L}}=\boldsymbol{I}-\boldsymbol{D^{-1}W}
\end{eqnarray}
being $\boldsymbol{W}$ the adjacency matrix of the network and $\boldsymbol{D}$ the diagonal matrix with entries $D_{kk}=s_{k}=\sum\limits_{i=1}^{N}W_{ki}$. 

The corresponding density matrix~\cite{de2016spectral} is defined by
\begin{eqnarray}
\boldsymbol{\rho}(\tau)=\frac{e^{-\tau \boldsymbol{\mathcal{L}}}}{\tr{e^{-\tau \boldsymbol{\mathcal{L}}}}}.
\end{eqnarray}

The spectral entropy of the network is therefore calculated according to the definition of Von Neumann entropy in case of a quantum system:
\begin{eqnarray}
S(\tau)&=& -\tr{\boldsymbol{\rho}(\tau)\log_{2}\boldsymbol{\rho}(\tau)}\nonumber\\
\label{eq:defentropy}
&=&\frac{1}{Z(\tau)\log 2}\tr{ \tau \boldsymbol{\mathcal{L}}  e^{-\tau \boldsymbol{\mathcal{L}}}} + \log_{2} Z(\tau), \nonumber
\end{eqnarray}
where $Z(\tau)=\tr{e^{-\tau \boldsymbol{\mathcal{L}}}}$ is the partition function.

In the following we will use the notation $\zint(\tau)$ and $\sint(\tau)$ to indicate, respectively, partition function and spectral entropy of a multiplex network. We will avoid to specify the index where this choice does not generate ambiguity.

If the multiplex consists of $L$ layers, we will use the notation $Z^{(\ell)}(\tau)$ and $S^{(\ell)}(\tau)$ for each layer $\ell=1,2,...,L$ separately.

\section{Random walk dynamics on multiplex} \label{app:randommultiplex}

Random walk dynamics on top of multiplex systems has not been yet defined. Therefore, for our purposes, we firstly introduce a more general framework to derive the multiplex master equation as a function of diffusion time $\tau$.

In fact, when exploring a layer $\ell$ in isolation, a random walker jumps from node $i$ to node $j$ with probability $T^{(\ell)}_{ij}=\frac{A^{(\ell)}_{ij}}{k^{(\ell)}_{i}}$, where $\mathbf{T}^{(\ell)}$ is the transition matrix of the random walk, $\mathbf{A}^{(\ell)}$ is the adjacency matrix of the layer and $k^{(\ell)}_{i}=\sum\limits_{j=1}^{N} A^{(\ell)}_{ij}$ is degree of node $i$ in layer $\ell$. 
Interacting layers in a multiplex network exhibit a richer dynamics, as the random walker in node $i$ can first switch from layer $\ell'$ to layer $\ell$ with probability $P^{(i)}_{\ell'\ell}$, and then jump from node $i$ to node $j$ with probability $T^{(\ell)}_{ij}$. Although all results of this paper are valid for any probability distribution, in absence of knowledge about the relative importance of weights,let us assume in the following that the probability to switch layer is uniform, i.e. $P_{\ell'\ell}=1/L$: then the overall probability to observe a transition from $i$ to $j$, regardless of the layer, is given by $T_{ij}=\frac{1}{L}\sum\limits_{\ell=1}^{L}T^{(\ell)}_{ij}$, which can be written in a more compact notation as $\mathbf{T}=\langle\mathbf{T}^{(\ell)}\rangle$. The weighted average, in case of further knowledge about the layer's weights, becomes: $\langle\mathbf{T}^{(\ell)}\rangle=\sum\limits_{\ell=1}^{L} P^{(\ell)} T^{(\ell)}$, where $P^{(\ell)}$ is the contribution of layer $\ell$ in the total flow.

It is worth remarking once again that the dynamics is dramatically different from aggregating all layers to a single network and then calculating the transition matrix of the random walk. In fact, two layers encode networks which are represented by adjacency matrices: aggregating two layers is achieved by summing entry-wise their adjacency matrices. Similarly, superimposing layers makes them indistinguishable to the eyes of random walker, so the dynamics can be seen as random walk on the aggregated network. Evidently, the distinguishability of layers' dynamics, plays an important role in transport phenomena of complex systems and cannot be simply neglected. In the following, we will use the terms aggregation and superposition interchangeably. Two or more layers are functionally grouped together, or aggregated, if the corresponding networks are superimposed.

Having the transition matrix, we can find the Laplacian matrix of multiplex given by $\mathcal{L}=\mathbf{I}-\mathbf{T}$. This allows for finding the master equation for random walk on multiplex networks. If the $i$--th components of the vector $\mathbf{p}(\tau)$ indicates the probability to find the random walker in node $i$ at time $\tau$, its evolution is governed by the master equation
\begin{eqnarray}
\mathbf{p}(\tau+1) = \mathbf{p}(\tau) \mathbf{T}
\end{eqnarray}
which, in the continuous-time approximation reduces to (Eq.~(\ref{eq:mastereq}).)

\begin{eqnarray}
\label{eq:mastereq}
\frac{\partial\mathbf{p}(\tau)}{\partial \tau} +\mathbf{p}(\tau) \boldsymbol{\mathcal{L}}  &=& 0, 
\end{eqnarray}
with solution given by $\mathbf{p}(\tau)=\mathbf{p}(0) e^{-\tau \boldsymbol{\mathcal{L}}}$. Here, $\boldsymbol{\mathcal{L}}$ indicates the normalized Laplacian matrix of the multiplex, $\mathbf{p}(\tau)$ is the vector encoding the probability of finding the random walker in a specific node at time $\tau$, and $e^{-\tau \boldsymbol{\mathcal{L}}}$ is the propagator of random walk dynamics. Note that the Laplacian matrix of the multiplex network is derived in terms of the average of Laplacian matrices $\boldsymbol{\mathcal{L}}^{(\ell)}$ $(\ell=1,2,...,L)$ of the single layers, i.e., $\boldsymbol{\mathcal{L}}=\langle\boldsymbol{\mathcal{L}^{(\ell)}}\rangle$.

\section{Dynamical trapping}\label{app:dynamicaltrapping}

To show this, let $\mathbf{p}(0)=\mathbf{p}(i|0)\equiv(0,...,1,0,...,0)$ indicate that the walk originates in node $i$ at time 0 with probability 1. In practice $\mathbf{p}(i|0)=\mathbf{e}_{i}$ is the $i$--th canonical vector in $\mathbb{R}^{N}$, being $N$ the size of the network. The \emph{return probability} for node $i$ at time $\tau$ is the probability of finding the random walker in $i$ at time $\tau$ steps later assuming it originated in $i$ at time 0, i.e. $\mathbf{p}(\tau)\mathbf{p}^{\dag}(i|0)$, where $^\dag$ indicates the transpose operator. From the solution of Eq.~(\ref{eq:mastereq}) in terms of the random walk propagator:
\begin{eqnarray}
\mathbf{p}(\tau) \mathbf{p}^{\dag}(i|0) = \mathbf{p}(0)e^{-\tau \boldsymbol{\mathcal{L}}} \mathbf{p}^{\dag}(0) = (e^{-\tau \boldsymbol{\mathcal{L}}})_{ii}.
\end{eqnarray}
Consequently, the \emph{average return probability} is given by 
\begin{eqnarray}
\mathcal{R}(\tau)=\frac{1}{N} \sum \limits_{i=1}^{N} (e^{-\tau \boldsymbol{\mathcal{L}}})_{ii} =\frac{1}{N} \tr{e^{-\tau \boldsymbol{\mathcal{L}}}}=\frac{Z(\tau)}{N}.
\end{eqnarray}
\section{Properties of the partition function}\label{app:partition}

Spectral entropy can be used to compare two networks~\cite{de2016spectral}. One measure of similarity is the Kullback-Leibler (KL) divergence. Spectral KL divergence is non-negative, exactly as its information-theoretic counterpart.

Given a multiplex network with density matrix $\boldsymbol{\rho}(\tau)$ and any of its layers, indicated by $l=1,2,...,L$, the following inequality holds:

\begin{eqnarray}
\mathcal{D}_{KL}(\boldsymbol{\rho}||\boldsymbol{\rho}^{(\ell)})=\tr{\boldsymbol{\rho}(\log_2 \boldsymbol{\rho} - \log_2 \boldsymbol{\rho}^{(\ell)})}\geq 0.
\end{eqnarray}
Note that the explicit dependence on $\tau$ is omitted for sake of clarity.

The average entropy distance of multiplex and its layers, can be quantified by the Kullbeck-Leibler entropy divergence, and in case of uniform distribution of layer weights, defines the intertwining as follows:
\begin{eqnarray}
\label{eq:avgdiv}
\mathcal{I} &=& \langle \mathcal{D}_{KL}(\boldsymbol{\rho}||\boldsymbol{\rho}^{(\ell)}) \rangle =\frac{1}{L} \sum\limits_{\ell=1}^{L} \mathcal{D}_{KL}(\boldsymbol{\rho}||\boldsymbol{\rho}^{(\ell)})\nonumber \\
&=&\frac{1}{L}\sum\limits_{\ell=1}^{L} \log\( Z^{(\ell)}\)-\log Z ,
\end{eqnarray}
which is zero if the system as a whole does not provide any additional information about its layers in isolation. Conversely, a large average divergence between the system and its layers highlights the necessity for using the multiplex description to encode the layer-layer interactions. In the next sections, we will show how the normalized version of this measure can be understood in terms of the entropy of layers in isolation and multiplex as a whole, and how it can be used to enhance transport properties while superimposing subsets of layers. We emphasize again that also intertwining can be found for any arbitrary set of layer weights, in case we have a way to asses the relative importance of layers.

Using Eq.~(\ref{eq:defentropy}) we obtain
\begin{eqnarray}
\tau \tr{\boldsymbol{ \mathcal{L}}^{(\ell)} \boldsymbol{\rho}} + \log Z^{(\ell)} - \tau \tr{ \boldsymbol{\mathcal{L} \rho}} - \log Z\geq0.
\end{eqnarray}
As it is true for all layers, we sum over layers and divide by the number of layers and exploit the relation $\boldsymbol{\mathcal{L}}=\frac{1}{L}\sum\limits_{\ell=1}^{L}\boldsymbol{\mathcal{L}}^{(\ell)}$ to obtain the average divergence
\begin{eqnarray}
\frac{1}{L} \sum\limits_{\ell=1}^{L} \mathcal{D}_{KL}(\boldsymbol{\rho}||\boldsymbol{\rho}^{(\ell)})=\log\(\prod\limits_{\ell=1}^{L} Z^{(\ell)}\)^{\frac{1}{L}}-\log Z \geq 0,\nonumber
\end{eqnarray}
from which the following fundamental inequality, relating the partition functions of layers with the one of the multiplex system, is obtained: 
\begin{eqnarray}
\label{eq:FR}
Z(\tau) \leq \prod\limits_{\ell=1}^{L} Z^{(\ell)}(\tau)^{\frac{1}{L}}.
\end{eqnarray}

A direct consequence of this result is that at least one layer $\ell^{\star}$ must have a partition function such that $Z^{(\ell^{\star})}(\tau)\geq Z(\tau)$, or inequality in Eq.~(\ref{eq:FR}) would not be satisfied. It follows:
\begin{eqnarray}
\label{ineq:logz}
\log Z^{(\ell^{\star})}(\tau) \geq \log Z(\tau),
\end{eqnarray}
and
\begin{eqnarray}
\label{ineq:zratio}
\tau \frac{Z^{(\ell^{\star})}(\tau)-1}{Z^{(\ell^{\star})}(\tau)} \geq \tau \frac{Z(\tau)-1}{Z(\tau)},
\end{eqnarray}
which will be useful in the following.

This last result holds for networks consisting of a single connected component, a scenario that could not be satisfied by many empirical systems. Its generalization to the case of multiple connected components is obtained as follows. First, let us assume that there are $C$ connected components in the system. If $\tau\gg 1$ then $Z(\tau) \approx C$ and the following approximation holds:
\begin{eqnarray}
\label{eq:logzapprox_0}
\log Z(\tau) = -\log\frac{1}{Z(\tau)} \approx \log C + \frac{Z(\tau)-C}{Z(\tau)},
\end{eqnarray}
leading to
\begin{eqnarray}
\label{ineq:zratio_2}
\tau\log C^{{(\ell^{\star})}} + \tau \frac{Z^{(\ell^\star)}(\tau)-C^{{(\ell^\star)}}}{Z^{(\ell^{\star})}(\tau)} \geq \tau\log C + \tau \frac{Z(\tau)-C}{Z(\tau)}. \nonumber
\end{eqnarray}

\section{Mean-field approximation}\label{app:mf}

The eigenvalue spectrum of the normalized Laplacian satisfies the following properties:
\begin{itemize}
\item $0 = \lambda_{1} \leq \lambda_{2} \leq ...\leq \lambda_{N}  \leq 2$;
\item $\tr{\boldsymbol{\mathcal{L}}}=\sum\limits_{i=1}^{N} \lambda_{i} = N $.
\end{itemize}

At this step, it is worth noting that $\boldsymbol{\rho}$ and $\boldsymbol{\mathcal{L}}$ can be eigen-decomposed as follows:
\begin{eqnarray}
\boldsymbol{\mathcal{L}}&=&\boldsymbol{Q\Lambda Q^{-1}},\\
\boldsymbol{\rho}(\tau)&=&\boldsymbol{Q}\frac{e^{-\tau\boldsymbol{\Lambda}}}{Z(\tau)}\boldsymbol{Q^{-1}},
\end{eqnarray}
being the columns of $\boldsymbol{Q}$ the eigenvectors of the normalized Laplacian matrix and $\boldsymbol{\Lambda}$ a diagonal matrix with $\Lambda_{ii}=\lambda_{i}$. It follows that $\nu_{i}(\tau)=e^{-\tau \lambda_{i}}/Z(\tau)$ and, as a consequence: 
\begin{eqnarray}
\tr{\boldsymbol{\mathcal{L} \rho}}= \sum\limits_{i=1}^{N} \lambda_{i} \nu_{i}(\tau)=\sum\limits_{i=C+1}^{N} \lambda_{i} \frac{e^{-\tau \lambda_{i}}}{Z(\tau)},
\end{eqnarray} 
the last step justified by the fact that $\lambda_1,...,\lambda_C=0$ for a network with $C$ connected components, which includes the possibility for $C_{\text{isol}}$ isolated nodes, i.e. nodes not interconnected with the rest of the system. For sake of completeness, it is worth remarking that $\tr{\boldsymbol{\mathcal{L}}}= N $ if there are multiple connected components but no isolated nodes, whereas $\tr{\boldsymbol{\mathcal{L}}} = N - C_{\text{isol}}$ if isolated nodes are present.

The mean-field approximation consists in neglecting higher-order terms in the following:
\begin{eqnarray}
\langle\lambda \nu(\tau)\rangle &=& \langle(\lambda -\bar{\lambda}+\bar{\lambda})(\nu(\tau)-\bar{\nu}(\tau)+\bar{\nu}(\tau))\rangle\nonumber\\
&=&\bar{\lambda}\bar{\nu}(\tau)+\langle(\lambda -\bar{\lambda})(\nu(\tau)-\bar{\nu}(\tau))\rangle \nonumber\\
&\approx&\bar{\lambda}\bar{\nu}(\tau).
\end{eqnarray}
To calculate the mean values, it is worth noting that:
\begin{eqnarray}
\bar{\lambda}=\frac{1}{N-C}\sum_{i=C+1}^{N}\lambda_{i}=\frac{N-C_{\text{isol}}}{N-C},
\end{eqnarray}
and
\begin{eqnarray}
\bar{\nu}(\tau)=\frac{1}{N-C}\sum_{i=C+1}^{N}\frac{e^{-\tau \lambda_{i}}}{Z(\tau)} = \frac{1}{N-C}\frac{Z(\tau)-C}{Z(\tau)}.\nonumber
\end{eqnarray}
It follows that
\begin{eqnarray}
\tr{\boldsymbol{\mathcal{L} \rho}}=(N-C)\langle\lambda \nu(\tau)\rangle\approx \frac{N-C_{\text{isol}}}{N-C}\frac{Z(\tau)-C}{Z(\tau)},\nonumber
\end{eqnarray}
which, for a network with no isolated nodes and only one connected component, it reduces to 
\begin{eqnarray}
\tr{\boldsymbol{\mathcal{L} \rho}}\approx\frac{Z(\tau)-1}{Z(\tau)},
\end{eqnarray}
for $N\gg 1$. It follows that in the thermodynamic limit the mean-field entropy is given by
\begin{eqnarray}
\smf^{\text{therm}}(\tau) = \frac{1}{\log 2} \( \tau \frac{Z(\tau)-1}{Z(\tau)} + \log Z(\tau) \),
\end{eqnarray}
or, more generally:
\begin{eqnarray}
\smf(\tau) = \frac{1}{\log 2} \( \tau \frac{N-C_{\text{isol}}}{N-C}\frac{Z(\tau)-C}{Z(\tau)} + \log Z(\tau) \), \nonumber
\end{eqnarray}
which, in the thermodynamic limit, reduces to
\begin{eqnarray}
\smf^{\text{therm}}(\tau) = \frac{1}{\log 2} \( \tau \frac{Z(\tau)-C}{Z(\tau)} + \log Z(\tau) \),
\end{eqnarray}
if $C$ scales sub-linearly with $N$, i.e., if the number of isolated nodes and disconnected components is much smaller than the size of the system, which might not be the case for finite empirical systems.\\
Taylor expanding the logarithm, keeping the first term in the limit of large $\tau$,
\begin{eqnarray}
\smf^{\text{therm}}(\tau) &=& \frac{1}{\log 2} \( \tau \frac{Z(\tau)-C}{Z(\tau)} + \log C + \frac{Z(\tau)-C}{Z(\tau)}\) \nonumber \\
&=& \frac{1}{\log 2} \( (\tau+1) \frac{Z(\tau)-C}{Z(\tau)} + \log C \) \nonumber \\
&=& \frac{(\tau+1)}{\log 2}   \frac{Z(\tau)-C}{Z(\tau)} + \log_{2} C,
\end{eqnarray}

\begin{figure}[!t]
\captionsetup[subfigure]{labelformat=empty}
\subfloat[]{\label{fig:bam5}%
  \includegraphics[width=0.25\textwidth]{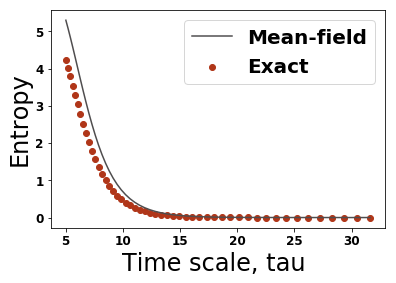}
}
\subfloat[]{\label{fig:bam8}%
  \includegraphics[width=0.25\textwidth]{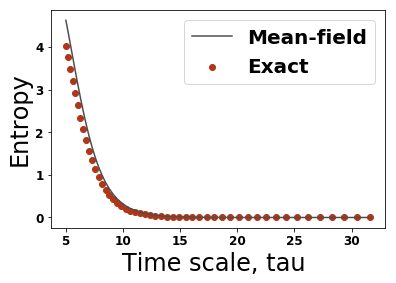}
}
\caption{\label{fig:MF_1}Entropy of Barabasi-Albert networks with $m=5$ (left) and $m=8$ (right), compared with the mean-field expectation.}
\end{figure}

\begin{figure}[!t]
\captionsetup[subfigure]{labelformat=empty}
\subfloat[$p=0.2$]{\label{fig:er20}%
  \includegraphics[width=0.25\textwidth]{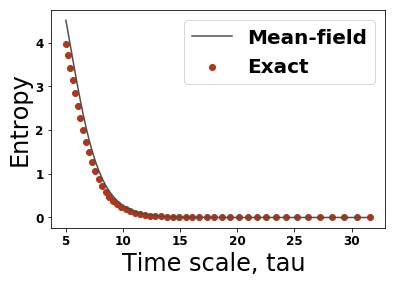}
}
\subfloat[$p=0.5$]{\label{fig:er50}%
  \includegraphics[width=0.25\textwidth]{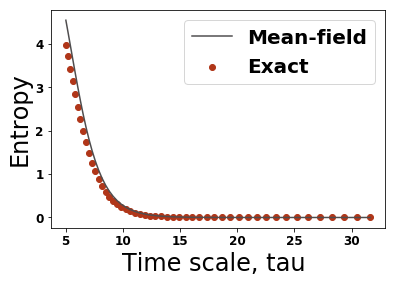}
}
\caption{\label{fig:MF_2}Entropy of Erdos-Renyi networks with $p=0.2$ (left) and $p=0.5$ (right), compared with the mean-field expectation.}
\end{figure}

The goodness of the mean-field approximation is numerically investigated for a variety of network models (see Fig.~\ref{fig:MF_1} and \ref{fig:MF_2} for emblematic examples).

\section{First order correction}\label{app:mf1st}

In the previous section we have neglected the cross-correlation term $\langle(\lambda -\bar{\lambda})(\nu(\tau)-\bar{\nu}(\tau))\rangle$ to obtain the mean-field approximation of the spectral entropy. It is straightforward to show that this term can be considered as a correction to the mean-field:
\begin{eqnarray}
\delta \smf(\tau)&=& \frac{\tau}{\log2} \langle(\lambda -\bar{\lambda})(\nu(\tau)-\bar{\nu}(\tau))\rangle\\\nonumber
&=&\frac{1}{\log2} \frac{\tau e^{-\tau}}{Z(\tau)}\sum_{i=2}^{N}\delta_{i}e^{-\delta_{i}\tau},
\end{eqnarray}

in order to write the spectral entropy as $S(\tau)=\smf(\tau)+\delta \smf(\tau)$. Note that $\delta_{i}=\lambda_{i}-\langle\lambda\rangle$, where $\lambda_{i}$ is the $i$-th eigenvalue of the normalized Laplacian matrix and $\langle\lambda\rangle=1$ in the thermodynamic limit.

In fact, $Z(\tau)\geq 1$ for any value of $\tau$, therefore the following inequality holds:
\begin{eqnarray}
\delta \smf(\tau)\leq -\tau^{2}e^{-\tau}(N-1)(\langle \lambda^2\rangle-1),
\end{eqnarray}
where we have expanded the exponential in its Taylor series truncated at second order and we have used the fact that $\langle \delta^2\rangle=\langle \lambda^2\rangle+1$. In the case of $C$ disconnected components, this difference is
\begin{eqnarray}
\delta \smf(\tau)\leq -\tau^{2}e^{-\tau}(N-C)(\langle \lambda^2\rangle-1).
\end{eqnarray}

From the definition of $\smean(\tau)$ and the above results, it is straightforward to show that
\begin{eqnarray}
\smean(\tau) &\leq& \frac{1}{L}\sum_{\ell=1}^{L}\smf^{(\ell)}(\tau) - \tau^{2}e^{-\tau}\frac{N-1}{L}\sum_{\ell=1}^{L}(\langle \lambda^{(\ell)^2}\rangle-1)\nonumber\\
\sint(\tau) &\leq& \sintmf(\tau) - \tau^{2}e^{-\tau}(N-1)(\langle \lambda_{\text{int}}^2\rangle-1). \nonumber
\end{eqnarray}
For a stochastic variable $X$ bounded in an interval $[x_{m},x_{M}]$ the Popoviciu's inequality holds:
\begin{eqnarray}
\text{Var}(X)\leq \frac{(x_M-x_m)^2}{4},
\end{eqnarray}
that in our case reduces to $\text{Var}(\lambda)\leq 1$, for the distribution of normalized Laplacian matrix's eigenvalues. It follows that $0\leq\langle \lambda^2 \rangle=\text{Var}(\lambda)+\langle\lambda\rangle^2\leq 2$. For $\tau\gg1$ then
\begin{eqnarray}
\smean(\tau) &\leq& \frac{1}{L}\sum_{\ell=1}^{L}\smf^{(\ell)}(\tau)\\
\sint(\tau) &\leq& \sintmf(\tau).
\end{eqnarray}

\section{Non-negativity of Intertwining}\label{app:intertwining}

From the fundamental inequality , it's straightforward to show that
\begin{eqnarray}
\label{ineq:1}
\log\zint(\tau) \leq \frac{1}{L} \sum_{\ell=1}^{L} \log Z^{(\ell)}(\tau).
\end{eqnarray}
If $\tau\gg 1$ then $Z(\tau) \approx 1$ and the following approximation holds:
\begin{eqnarray}
\label{eq:logzapprox}
\log Z(\tau) = -\log\frac{1}{Z(\tau)} \approx \frac{Z(\tau)-1}{Z(\tau)},
\end{eqnarray}
which leads to
\begin{eqnarray}
\label{ineq:2}
\frac{\zint(\tau)-1}{\zint(\tau)} \leq \frac{1}{L} \sum_{\ell=1}^{L} \frac{Z^{(\ell)}(\tau)-1}{Z^{(\ell)}(\tau)}.
\end{eqnarray}
By summing side by side inequalities (\ref{ineq:1}) and (\ref{ineq:2}) we obtain
\begin{eqnarray}
\sint(\tau) \leq \smean(\tau),
\end{eqnarray}
leading to non-negativity of the intertwining
\begin{eqnarray}
\mathcal{I}(\tau)=\smean(\tau)-\sint(\tau)\geq 0,
\end{eqnarray}
for $\tau\gg1$. We wonder to which extent this inequality holds for smaller values of $\tau$.

After expanding $Z$ around  $\delta_{i}=\lambda_{i}-\langle \lambda \rangle = \lambda_{i} - 1 $  we have at first-order:
\begin{eqnarray}
Z(\tau)&=& 1 + \sum_{i=2}^{N} e^{-\tau \lambda_{i}} = 1 + e^{-\tau} \sum_{i=2}^{N} e^{-\tau \delta_{i}} \nonumber\\
&=& 1 +  (N-1) e^{-\tau} \langle 1 - \tau \delta + \frac{\tau^{2}}{2} \delta^{2} - ... \rangle \nonumber \\ 
&\approx& 1 + (N-1) e^{-\tau}, 
\end{eqnarray}
from which $Z^{-1}(\tau) \approx 1 - (N-1) e^{-\tau}$. 

It follows that Eq.~(\ref{eq:logzapprox}) is satisfied provided that $(N-1)e^{-\tau} \ll 1$, which is equivalent to require that $\tau$ should be larger than the $\tau_{c}$ where $\tau_{c} > \log(N-1)$ in order for mean-field approximation to be very accurate.

In the more general case, we have $Z(\tau) \approx C + (N-C) e^{-\tau}$, from which 
\begin{eqnarray}
Z^{-1}(\tau) \approx C^{-1}(1 - \frac{N-C}{C} e^{-\tau}), 
\end{eqnarray}
which similarly holds if $\tau_{c} > \log(\frac{N}{C}-1)$ so that $\log\frac{Z}{C}$ can be expanded in the same way. Note that if there are isolated nodes, the fundamental inequality must be proven.

\section{When and why the structural reducibility fails}\label{app:classicalfail}

Of course, the non-interacting scenario described here is barely observed in reality and $\snint(\tau)$ would be equal to the entropy of one single layer. However, our framework suggests that even in the case of different layers, we can interpret $\snint(\tau)$ as an upper bound -- obtained by exploring the system using each layer separately -- to $\sint(\tau)$ (see Appendices~\ref{app:mf}, \ref{app:mf1st} and \ref{app:intertwining}). Then, any deviation from this upper bound highlights the existence of layer-layer interactions and can be used functional reduciblity of the system. In the light of this distinction, we are now able to better understand existing measures in the literature. For instance, the classical structural reducibility~\cite{de2015structural} is based on comparing $\snint$ against the entropy of the fully aggregated system $\sagg$: clearly this method has to fail when some layers are exactly the same and, more generally, when they are very similar with each other (Appendix~\ref{app:classicalfail}). 

The classical structural reducibility method is based on maximizing the quality function 
\begin{eqnarray}
q(n)=1-\frac{\langle S^{(\ell)}\rangle(n)}{S_{\text{agg}}},
\end{eqnarray}
where $n$ is the aggregation step, and $S_{agg}$ is the entropy of the fully aggregated system.   
Let us consider a multiplex network with $L$ layers, where $m$ layers are identical and have entropy $S^{(\text{same})}=S^{(\ell)}$ $(\ell = 1, 2, ... m)$, while the remaining layers have different topologies and we indicate by 
\begin{eqnarray}
S^{(dif)}=\frac{1}{L-m}\sum\limits_{\ell=L-m+1}^{L}S^{(\ell)}
\end{eqnarray}
their average entropy. We expect a suitable optimization procedure to identify the $m-1$ redundant layers and reduce the multiplex system to $L-m+1$ layers. 

During the first $m$ aggregation steps, the layers that are supposed to be aggregated are exactly the same. Remarking that the entropy remains unchanged if the adjacency matrix is multiplied by a constant, when two out of $m$ equal layers are aggregated, the entropy of the resulting network is still $S^{(same)}$. So when calculating the average entropy of layers, aggregation of every pair of equal layers is exactly equal to removing one of them. This way it is straightforward to write the average entropy of layers as a function of aggregation step $n$ which follows,
\begin{eqnarray}
\langle S^{(\ell)} \rangle (n)&=& \frac{1}{L-n} \sum\limits_{\ell =n+1}^{m} S^{(\ell)} \nonumber \\ 
&=&\frac{1}{L-n} \[\sum\limits_{\ell =n+1}^{m} S^{(\ell)} +  \sum \limits_{\ell=L-m+1}^{L} S^{(\ell)} \] \nonumber \\ 
&=&\frac{m-n}{L-n} S^{(same)} + \frac{L-m}{L-n} S^{(dif)} \nonumber \\
&=& \frac{1}{L-n} \[(m-n) S^{(same)} + (L-m) S^{(dif)} \] \nonumber
\end{eqnarray} 
Writing one of the entropies in terms of the other one $ S^{(dif)} = \delta + S^{(same)}  $ we have
\begin{eqnarray}
\langle S^{(\ell)} \rangle (n) &=& \frac{1}{L-n} [(m-n) S^{(same)} \nonumber \\
&+& (L-m) S^{(same)} + (L-m) \delta ] \nonumber \\
&=& S^{(same)} + \frac{L-m}{L-n} \delta 
\end{eqnarray}

\begin{figure}[!ht]
\centering
\includegraphics[width=0.45\textwidth]{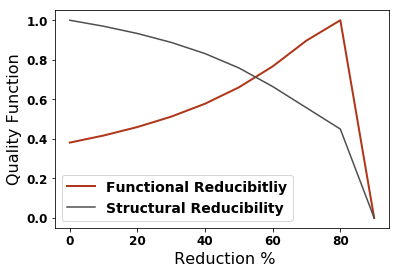}
\caption{\label{fig:Last_method}Functional versus structural redundancy. A multiplex network with 10 layers is given, in which 9 layers are exactly the same and expected to be grouped into one network. In scenarios like this one, the structural reducibility fails to identify the redundant layers to superimpose. The curve corresponding to functional redundancy has been rescaled to match the scale of the other curve to facilitate the comparison of trends and the identification of maxima.}
\end{figure}


Therefore, if $S^{(dif)}>S^{(same)}$ then $\delta$ is positive, and the aggregation of similar layers increases the average entropy of layers $\langle S^{(\ell)}\rangle(n)$, while decreasing the classical quality function: $q(n)=1-\frac{\langle S^{(\ell)}\rangle(n)}{S_{agg}}$. This case corresponds to an irreducible structure, an evident failure of the method because it is only comparing the average entropy of different groups of layers. A convincing example is given in Fig.~\ref{fig:Last_method}, where the method is applied to a multiplex system consisting of 10 layers (9 are identical).

\section{Using intertwining for dimensionality reduction}\label{app:intertwining_dimred}

If $\boldsymbol{\mathcal{L}}$ indicates the normalized Laplacian matrix of the multiplex network, let us introduce perturbation matrices $\Delta\boldsymbol{\mathcal{L}}^{(\ell)}$ such that the normalized Laplacian matrix of each layer can be written as 
\begin{eqnarray}
\boldsymbol{\mathcal{L}}^{(\ell)} = \boldsymbol{\mathcal{L}} + \Delta\boldsymbol{\mathcal{L}}^{(\ell)}.
\end{eqnarray}
What is the effect of such perturbations on the corresponding partition functions? In fact, we have already shown that
\begin{eqnarray}
Z(\tau)\simeq 1+(N-1)e^{-\tau}[1+\frac{\tau^2}{2}(\langle\lambda^{2}\rangle-1)  - ...]
\end{eqnarray}
either for the multiplex network or for each layer separately. This result suggests that it is sufficient to find a relation between the second-order moments of the eigenvalue distribution for networks separately and the whole multiple system.

In the following, we use the perturbation approach suggested by Lord Rayleigh, where we implicitly assume that perturbations to the Laplacian matrices are small. Since matrices $\boldsymbol{\mathcal{L}}^{(\ell)}$ are diagonalizable, then
\begin{eqnarray}
\lambda_{i}^{(\ell)}=\lambda_{i} + \Delta\lambda_{i}^{(\ell)},
\end{eqnarray}
with
\begin{eqnarray}
\Delta\lambda_{i}^{(\ell)}=\mathbf{q}_{i}^{\top}\Delta \boldsymbol{\mathcal{L}}^{(\ell)} \mathbf{q}_{i},
\end{eqnarray}
where $\mathbf{q}_{i}$ indicates the $i$-th eigenvector of $\boldsymbol{\mathcal{L}}$. Since $\langle\lambda^{(\ell)}\rangle=1$ and $\langle\lambda\rangle=1$, it follows that $\langle\Delta\lambda^{(\ell)}\rangle=0$, whereas
\begin{eqnarray}
\left\langle\(\lambda^{(\ell)}\)^{2}\right\rangle = \left\langle\lambda^{2}\right\rangle + \left\langle\(\Delta\lambda^{(\ell)}\)^{2}\right\rangle,
\end{eqnarray}
where the covariance term $\left\langle\lambda\Delta\lambda^{(\ell)}\right\rangle=0$ because perturbed eigenvalues are not depending from the eigenvalues of $\boldsymbol{\mathcal{L}}$. It follows that
\begin{eqnarray}
Z^{(\ell)}(\tau)=\zint(\tau)+\Delta Z^{(\ell)}(\tau),
\end{eqnarray}
with 
\begin{eqnarray}
\Delta Z^{(\ell)}(\tau)=\frac{\tau^{2}}{2}e^{-\tau}(N-1)\left\langle\(\Delta\lambda^{(\ell)}\)^{2}\right\rangle.
\end{eqnarray}

We can also write the product of partition functions of layers in terms of perturbations:
\begin{eqnarray}
\prod_{\ell=1}^{L} Z^{(\ell)^\frac{1}{L}}(\tau) &=& \prod_{\ell=1}^{L} \(\zint(\tau) + \Delta Z^{(\ell)}(\tau)\)^{\frac{1}{L}}\\
&=& \zint(\tau) \prod_{\ell=1}^{L}  \( 1 + \frac{\Delta Z^{(\ell)}(\tau)}{\zint(\tau)}\)^{\frac{1}{L}}.
\end{eqnarray}
Since $1 \gg \frac{\Delta Z^{(\ell)}(\tau)}{\zint(\tau)}$ we can use the approximation $\( 1 + \frac{\Delta Z^{(\ell)}(\tau)}{\zint(\tau)}\)^{\frac{1}{L}} \approx 1 + \frac{1}{L} \frac{\Delta Z^{(\ell)}(\tau)}{\zint(\tau)}$ at first order, to obtain 
\begin{eqnarray}
\prod_{\ell=1}^{L}  \( 1 + \frac{\Delta Z^{(\ell)}(\tau)}{\zint(\tau)}\)^{\frac{1}{L}} 
&\approx&\prod_{\ell=1}^{L}  \( 1 + \frac{1}{L} \frac{\Delta Z^{(\ell)}(\tau)}{\zint(\tau)}\) \nonumber\\ 
&=&  1 + \frac{1}{L} \sum_{\ell=1}^{L} \frac{\Delta Z^{(\ell)}(\tau)}{\zint(\tau)} \nonumber\\
&+& \mathcal{O}\(\Delta Z^{(\ell)^2}(\tau) , ... \) \nonumber\\ 
&\approx& 1 + \overline{\Delta Z^{(\ell)}}(\tau), 
\end{eqnarray}
where
\begin{eqnarray}
\overline{\Delta Z^{(\ell)}}(\tau) &\approx& \frac{\tau^{2}}{2}e^{-\tau}(N-1)\overline{\left\langle\(\Delta\lambda^{(\ell)}\)^{2} \right\rangle} \nonumber
\end{eqnarray}
and
\begin{eqnarray}
\overline{\left\langle\(\Delta\lambda^{(\ell)}\)^{2} \right\rangle}=\frac{1}{L} \sum_{\ell=1}^{L} \left\langle\(\Delta\lambda^{(\ell)}\)^{2} \right\rangle \nonumber
\end{eqnarray}
Above the time $\tau_{c}$, Eq.~(\ref{eq:logzapprox}) holds and we can write entropy in mean-field regime as
\begin{eqnarray}
\smf(\tau) \times \log 2 &=& \tau \frac{Z(\tau)-1}{Z(\tau)} + \log Z(\tau) \nonumber\\
&\approx& (\tau+1) \frac{Z(\tau)-1}{Z(\tau)}.
\end{eqnarray}

On the one hand, let us consider the average information divergence given by
\begin{eqnarray}
\frac{1}{L} \sum\limits_{\ell=1}^{L} \mathcal{D}_{KL}(\boldsymbol{\rho}||\boldsymbol{\rho}^{(\ell)})&=& 
\frac{1}{L}\sum\limits_{\ell=1}^{L} \log\( Z^{(\ell)}\)-\log Z,\nonumber
\end{eqnarray}
where, multiplying by $\tau+1$ we obtain the important relation with the entropy deviation given by
\begin{eqnarray}
(\tau+1)\frac{1}{L} \sum\limits_{\ell=1}^{L}
\mathcal{D}_{KL}(\boldsymbol{\rho}||\boldsymbol{\rho}^{(\ell)})&=& 
(\tau+1)\frac{1}{L}\sum\limits_{\ell=1}^{L} \log\( Z^{(\ell)}\) \nonumber\\
&-&(\tau+1)\log Z \nonumber\\
&=& \smean(\tau) - \sint(\tau). \nonumber
\end{eqnarray}

On the other hand, since
\begin{eqnarray}
\log{\frac{\prod\limits_{\ell=1}^{L} Z^{(\ell)^\frac{1}{L}}(\tau)}{\zint(\tau)}} &=& \log \(1+\frac{\overline{\Delta Z^{(\ell)}}(\tau) }{\zint(\tau) }\) \approx \frac{ \overline{\Delta Z^{(\ell)}}(\tau) }{\zint(\tau)},\nonumber
\end{eqnarray}
we can write
\begin{eqnarray}
(\tau + 1)\log{\frac{\prod \limits_{\ell=1}^{L} Z^{(\ell)}(\tau)^{\frac{1}{L}}}{\zint(\tau)}}&=& (\tau+1) \frac{1}{L} \sum_{\ell=1}^{L} \frac{Z^{(\ell)}(\tau)-1}{Z^{(\ell)}(\tau)}\nonumber\\ 
&-& (\tau+1) \frac{\zint(\tau)-1}{\zint(\tau)},\nonumber
\end{eqnarray}
from which we obtain
\begin{eqnarray}
\smean(\tau) - \sint(\tau)=\mathcal{I}(\tau) &\approx& (\tau+1) \frac{\overline{\Delta Z^{(\ell)}}(\tau)}{\zint(\tau)}.\nonumber
\end{eqnarray}
It is worth remarking the dependence of the intertwining on the average variance of single-layer eigenvalues, stressing the fact that it can be used as a measure of topological diversity of layers. In practice, it is desirable to work with the relative intertwining, defined by
\begin{eqnarray}
\mathcal{I}^\star(\tau) = 1 - \frac{\sint(\tau)}{\smean(\tau)}.\nonumber
\end{eqnarray}
Since $\sint(\tau)\leq \smean(\tau)$ then $0\leq \smean(\tau)-\sint(\tau)\leq \smean(\tau)$, therefore $0\leq \mathcal{I}^\star(\tau)\leq 1$. For small perturbations, the relative intertwining reduces to 
\begin{eqnarray}
\mathcal{I}^{\star}(\tau)&=&\frac{(\tau+1)\frac{\overline{\Delta Z^{(\ell)}}(\tau)}{\zint(\tau)} }{(\tau+1)\frac{1}{L}\sum\limits_{\ell=1}^{L}\frac{ Z^{(\ell)}(\tau) -1}{ Z^{(\ell)}(\tau)}}  = \frac{\frac{\overline{\Delta Z^{(\ell)}}(\tau)}{\zint(\tau)} }{\frac{1}{L}\sum\limits_{\ell=1}^{L}\frac{\zint(\tau)+ \Delta Z^{(\ell)}(\tau) -1}{\zint(\tau)+\Delta Z^{(\ell)}(\tau)}}   \nonumber \\
&\approx& \frac{\frac{\overline{\Delta Z^{(\ell)}}(\tau)}{\zint(\tau)} }{\frac{1}{L}\sum\limits_{\ell=1}^{L}\frac{\zint(\tau)+ \Delta Z^{(\ell)}(\tau) -1}{\zint(\tau)}(1 - \frac{\Delta Z^{(\ell)}(\tau)}{\zint})} 
\nonumber \\
&\approx& \frac{\frac{\overline{\Delta Z^{(\ell)}}(\tau)}{\zint(\tau)} }{\frac{1}{L}\sum\limits_{\ell=1}^{L}\frac{\zint(\tau)+ \Delta Z^{(\ell)}(\tau) -1}{\zint(\tau)}} = 
\frac{\overline{\Delta Z^{(\ell)}}(\tau)}{ \zint(\tau) +\overline{\Delta Z^{(\ell)}}(\tau) -1 } \nonumber \\ 
 &=& \frac{(N-1)  \frac{\tau^{2} e^{-\tau} }{2} \overline{\left\langle\(\Delta\lambda^{(\ell)}\)^{2} \right\rangle}}{(N-1)e^{-\tau} \left[1+\frac{\tau^2}{2}(\langle\lambda^{2}\rangle-1) +\frac{\tau^2}{2}\overline{\left\langle\(\Delta\lambda^{(\ell)}\)^{2} \right\rangle}\right]} \nonumber \\
 &=& \frac{ \frac{\tau^{2}}{2} \overline{\left\langle\(\Delta\lambda^{(\ell)}\)^{2} \right\rangle}}{ 1+\frac{\tau^2}{2}(\langle\lambda^{2}\rangle-1) + \frac{\tau^{2}}{2} \overline{\left\langle\(\Delta\lambda^{(\ell)}\)^{2} \right\rangle}} 
\end{eqnarray}
It is worth noting that the second-order approximation used in this section requires that $ \frac{\tau^{2}}{2} \overline{\left\langle\(\Delta\lambda^{(\ell)}\)^{2} \right\rangle}\ll 1$ and $\frac{\tau^2}{2}(\langle\lambda^{2}\rangle-1)\ll 1$. It follows that the relative intertwining can be well approximated by 
\begin{eqnarray}
\mathcal{I}^\star(\tau) &\approx& \frac{\tau^{2}}{2} \overline{\left\langle\(\Delta\lambda^{(\ell)}\)^{2} \right\rangle}.
\end{eqnarray}

The last relation highlights how the intertwining is expected to increase for increasing deviation of single-layer eigenvalue spectra from the overall average. Additionally, the relative intertwining can be approximated as $\mathcal{I}^{*}(\tau) \approx \frac{\overline{\Delta Z^{(\ell)}}(\tau)}{ \zint(\tau) +\overline{\Delta Z^{(\ell)}}(\tau) -1 }$. If the perturbation is sufficiently small, i.e. $\overline{\Delta Z^{(\ell)}}(\tau) \ll  \zint(\tau) $, the approximation reduces to $\mathcal{I}^{*}(\tau) \approx \frac{\overline{\Delta Z^{(\ell)}}(\tau)}{ \zint(\tau) -1 }$, which directly relates the intertwining to the partition function of multiplex network.

The integral over time of the relative intertwining is used as a quality function in the main text in order to have a parameter-independent quantification of diversity. If $m=0, 1, ..., L-1$ indicates the aggregation step of the functional reduction procedure, we define the quality function by
\begin{eqnarray}
\mathcal{I}^{\star}(m)=\frac{1}{\tau_{max}-\tau_{c}}\int_{\tau_{c}}^{\tau_{max}}d\tau~\mathcal{I}^\star(m;\tau),
\end{eqnarray}
where $\tau_{c}> \log(N-1)$ and $\tau_{max}$ can be arbitrarily large. To allow for a meaningful comparison with some transport properties, as shown in the next section, we can set $\tau_{max}=N$ without loss of generality: this choice corresponds to let random walkers explore the network with a number of steps that is of the order of system's size. Since $\mathcal{I}^\star(m,\tau)=f(\tau)\phi(m)$, i.e. its dependence on variables can be separated, the resulting quality function will not be affected by the integral of $f(\tau)$ at different aggregation steps, therefore we can focus only on the influence of $\phi(m)=\overline{\left\langle\(\Delta\lambda^{(\ell)}\)^{2} \right\rangle}$.

To better understand the behavior of $\phi(m)$, let us consider the case of a multiplex network consisting of $L$ layers, among which $m_{0}<L$ are identical. For sake of simplicity, but without loss of generality, we can assume that the identical layers are the ones labeled by $1,2,...,m_{0}$, where non-identical layers are labeled by $m_{0}+1, ..., L$. It follows that
\begin{eqnarray}
\phi(m)&=&\frac{m_{0}-m}{L-m}\left\langle\(\Delta\lambda^{(1)}\)^{2} \right\rangle \nonumber \\
&+& \frac{1}{L-m}\sum_{\ell=m_{0}+1}^{L}\left\langle\(\Delta\lambda^{(\ell)}\)^{2} \right\rangle \nonumber
\end{eqnarray}
and
\begin{eqnarray}
\phi(m+1)&-&\phi(m)=\frac{m_{0}-L}{(L-m)(L-m-1)}\left\langle\(\Delta\lambda^{(1)}\)^{2} \right\rangle \nonumber\\
&& + \frac{1}{(L-m)(L-m-1)}\sum_{\ell=m_{0}+1}^{L}\left\langle\(\Delta\lambda^{(\ell)}\)^{2} \right\rangle,\nonumber
\end{eqnarray}
suggesting that the difference in the quality function between two successive aggregation steps depends on the trade-off between a negative term (the first one, since $m_{0}<L$) and a positive term (the second one). This difference is positive if and only if
\begin{eqnarray}
\frac{1}{L-m_{0}}\sum_{\ell=m_{0}+1}^{L}\left\langle\(\Delta\lambda^{(\ell)}\)^{2} \right\rangle \geq \left\langle\(\Delta\lambda^{(1)}\)^{2} \right\rangle. 
\end{eqnarray}
The last result provides us with a deep insight on the behavior of the quality function: it increases after aggregating two layers that are identical -- or, equivalently, with an eigenvalue spectrum whose variance is identically displaced from the overall average -- if and only if the average variance of the eigenvalue spectra of non-identical layers is larger than the variance of the eigenvalue spectra of identical ones. Roughly speaking, the quality function increases when \revv{superimposing} two layers which provide redundant information in the multiplex system whereas it decreases when it is not the case.

Since at the latest possible aggregation step, i.e. $m=L-1$, the intertwining is zero by construction, when redundant layers are present we expect an optimal step $m_{opt}$ to exist in correspondence of the step where topological diversity is maximum.

\section{Collective cosine distance between layers}\label{app:CCD}

A standard way to quantify dissimilarity between layers is to use Jensen-Shannon entropy divergence. Indicating by $\boldsymbol{\rho}^{(\ell)}(\tau)$ and $\boldsymbol{\rho}^{(\ell')}(\tau)$ be the density matrices corresponding to two layers $\ell$ and $\ell'$ of a multiplex network, their Jensen-Shannon divergence is defined by
\begin{eqnarray}
\mathcal{J}[\boldsymbol{\rho}^{(\ell)}(\tau)||\boldsymbol{\rho}^{(\ell')}(\tau)]=S^{(\text{mix})}(\tau) - \frac{1}{2}[S^{(\ell)}(\tau)+S^{(\ell')}(\tau)],\nonumber
\end{eqnarray}
formally equivalent to the measure widely used in quantum computing. Here, $S^{(\text{mix})}$ is the entropy of the mixture matrix defined by $\boldsymbol{\mu}^{(\ell,\ell')}=[\boldsymbol{\rho}^{(\ell)}(\tau)+\boldsymbol{\rho}^{(\ell')}(\tau)]/2$. The quantity $\sqrt{\mathcal{J}[\boldsymbol{\rho}^{(\ell)}(\tau)||\boldsymbol{\rho}^{(\ell')}(\tau)]}$ defines a metric distance between layers that is a function of diffusion time $\tau$. 

However, Jensen-Shannon distance (JSD) might be not reliable in the case of multiplex systems consisting of multiple connected components and isolated nodes across layers, scenarios typical when analyzing empirical systems. Moreover, it depends on $\tau$: for multi-resolution analysis this is desirable, while it is less desirable when one is not interested in having results at different time scales.

In this section, we introduce a novel distance measure, that we name collective cosine distance (CCD), which is not affected by the above limitations. At variance with JSD, which is an information-theoretic measure, CCD is a geometric measure based on the calculation of the average angular distance between transition vectors. The transition vector $\vec{T}_{i}$ associated to node $i$ is a column of the transition matrix $\mathbf{T}$, and encodes the probability of jumps from that specific node to any other node in the network. As the role of a certain node in distribution of flow can vary from layer to layer, $\vec{T}_{i}^{(\ell)}$ and $\vec{T}_{i}^{(\ell')}$ ($\ell, \ell'=1,2,...,L$) can in general be different. Here, we quantify this difference by their angular distance in the Euclidean space $\mathbb{R}^{N}$. This distance is calculated as
\begin{eqnarray}
\theta_{i}^{(\ell,\ell')}= \cos^{-1}\left(\frac{\vec{T}_{i}^{(\ell)}\cdot\vec{T}_{i}^{(\ell')}}{|\vec{T}_{i}^{(\ell)}||\vec{T}_{i}^{(\ell')}|}\right),
\end{eqnarray}
where $|\cdot|$ denotes the vector norm. A collective variable of interest is the average angular distance of nodes across two arbitrary layers: 
\begin{eqnarray}
\Theta^{(\ell,\ell')}=\frac{1}{N}\sum\limits_{i=1}^{N}\theta_{i}^{(\ell,\ell')}.
\end{eqnarray}
 Note that in case of isolated state nodes, the corresponding transition vector can be taken as zero. The  collective cosine distance (CCD) between two layers, accounting for the collective angle, is defined by 
\begin{eqnarray}
d(\ell, \ell')=1-\cos(\Theta^{(\ell,\ell')})
\end{eqnarray} 
 and can be used to quantify the pairwise dissimilarity of layers while accounting for dissimilarities of their diffusion pathways.

It is worth remarking that CCD is more convenient than JSD for the goal of this study. From a computational perspective, it is dramatically faster than JSD and avoid hierarchical clustering: therefore, reduction is not based on a heuristics but it can be the result of an exhaustive search through all possible functional couplings. From a theoretical perspective, CCD does not depend on a temporal scale and naturally deals with the presence of isolated nodes and multiple connected components.

\section{Laplacian matrix in presence of isolated nodes}

Empirical multiplex systems usually consist of several isolated state nodes and, sometimes, of disconnected components. The presence of these isolated nodes leads to singularities when modeling the dynamics on top of the network. To avoid such singularities, a typical workaround is to add self-loops connecting isolated nodes to themselves. However, this mathematical trick boosts the dynamical trapping and consequently, it dramatically alters the resulting dynamics in an unpredictable way. Another possibility is to add virtual links, with extremely small weights, connecting an isolated state node to all state nodes of the corresponding layer, or to a number of them. This workaround provides a more realistic approach to model a diffusion process on top of multiplex networks. Unfortunately, when the number of isolated state nodes is large, a large portion of the flow goes through the virtual links and, consequently, the resulting dynamics in an unpredictable way.

To overcome those issues, we propose an alternative approach to find the right Laplacian matrix of a multiplex network and its layers when an arbitrary number of isolated state nodes is present. To this aim, we first question the physical nature of isolated state nodes. As they have no contribution in any interaction, they should not have any impact on the overall flow as well. More technically, instead of focusing on the structure, we associate suitable transition vectors to the isolated state nodes in order to remove their influence on the overall dynamics. 

Mathematically, let us indicate by $\vec{T}_{i}^{(\ell)}$ the transition vector for node $i$ in layer $\ell$, and, if the node is isolated in that layer, let us indicate the vector by $\vec{T}_{i}^{(\ell,iso)}$. Our goal is to exclude the effects of isolated state nodes on the transition vector of node $i$. Let us assume that node $i$ is connected in only $m$ out of $L$ layers of the multiplex network: the corresponding multiplex transition vector is given by 
\begin{eqnarray}
\vec{T}_{i}&=&\frac{1}{L}\sum\limits_{\ell=1}^{L} T_{i}^{(\ell)}= \frac{1}{L}\left[\sum\limits_{\ell=1}^{m} T_{i}^{(\ell)} + \sum\limits_{\ell= m+1 }^{L} T_{i}^{(\ell,iso)}\right]\nonumber\\
&=&\frac{1}{m}\sum\limits_{\ell=1}^{m} T_{i}^{(\ell)}, \nonumber
\end{eqnarray}
from which
\begin{eqnarray}
\frac{1}{L} \sum\limits_{i=m+1}^{L} \vec{T}_{i}^{(\ell,iso)} &=&  \frac{L-m}{mL}  \sum\limits_{\ell=1}^{m} \vec{T}_{i}^{(\ell)} . \nonumber
\end{eqnarray}

Assuming the same contribution for all isolated state nodes in their average, $\vec{T}_{i}^{(\ell,iso)}=\vec{T}_{i}^{(\ell',iso)}$, one obtains
\begin{eqnarray}
\frac{L-m}{L}\vec{T}_{i}^{(\ell,iso)}=\frac{L-m}{mL}\sum\limits_{\ell=1}^{m}\vec{T}_{i}^{(\ell)},
\end{eqnarray}{}
that simply reads
\begin{eqnarray}
\vec{T}_{i}^{(\ell,iso)}=\frac{1}{m}\sum\limits_{\ell=1}^{m}\vec{T}_{i}^{(\ell)}.
\end{eqnarray}

Once transition vectors for isolated state nodes are obtained, one can find the transition matrix of layers with isolated nodes, and consequently the transition matrix of the multiplex, as the weighted average of transition matrices of layers.

\section{Transport metrics}\label{app:transport}

In this section, we introduce some transport metrics that we used in numerical experiments to validate how functional reducibility enhances the transport properties in complex systems.

\heading{Diffusion time} The eigendecomposition of the propagator reads $e^{-\tau \boldsymbol{\mathcal{L}}}=\sum\limits_{i=1}^{N} \mathbf{C}(i) e^{-\lambda_{i} \tau}$. The evolution of the probability vector is then
\begin{eqnarray}
\mathbf{p}(\tau)=\sum\limits_{i=1}^{N}\mathbf{p}(0) \mathbf{C}(i) e^{-\tau \lambda_{i}}.
\end{eqnarray}
Since $\lambda_{1}=0$, the first term of the summation is a constant, whereas the other terms decay exponentially fast. The second smallest eigenvalue $\lambda_{2}$ governs the convergence of the process by setting a typical time scale $\lambda_{2}^{-1}$, which is known as \emph{diffusion time}.

\heading{Average return probability} The average return probability, within a time $\tau$, encodes the probability to return back to the origin of a random walk. It is is given by
\begin{eqnarray}
\mathcal{R}(\tau)=\int_{0}^{\infty}d\lambda~e^{-\tau \lambda}\rho(\lambda),
\end{eqnarray}
where $\rho(\lambda)$ is the spectral density of the normalized Laplacian matrix~\cite{barrat2008dynamical}. 

In order to allow for comparing different networks regardless of a specific temporal scale, we integrate again within the same interval $\Delta \tau=\tau_{f}-\tau_{i}$ used for the overall analysis. Finally, we use
\begin{eqnarray}
\mathcal{R}=\frac{1}{\tau_{f}-\tau_{i}} \int_{\tau_{i}}^{\tau_{f}}  d\tau~ \mathcal{R}(\tau).
\end{eqnarray}

Note that we perform the same integration for the partition function, to get rid of its dependence on $\tau$.

\heading{Navigability} One way to define the \emph{navigability} of a multiplex network is to measure its coverage~\cite{de2014navigability}, quantifying the efficiency of the walker in discovering the network in terms of the average fraction of distinct nodes visited at least once within $\tau$ steps. For connected networks, this fraction increases over time until the random walker explores all nodes of the network. A good approximation for the coverage is given by,
\begin{eqnarray}
\Omega(\tau) \approx 1 - \frac{1}{N^{2}} \sum\limits^{N}_{i,j=1} (1-\delta_{ij}) e^{-C_{ij}(1) \tau -C_{ij}(2)\lambda_{2}^{-1}},
\end{eqnarray}
where $\delta_{ij}$ is the Kronecker delta, $\lambda_{2}$ is the second smallest eigenvalue and $\mathbf{C}(i)$ depends on the $i$--th eigenvector of the normalized Laplacian matrix. To compare two networks in terms of their coverage, we define their navigability by integrating the corresponding coverage function over time and then normalizing it by the integral vicinity:
\begin{eqnarray}
\mathcal{N} = \frac{1}{\tau_{f}-\tau_{i}} \int\limits_{\tau_{i}}^{\tau_{f}}  d\tau \Omega(\tau).
\end{eqnarray}

\begin{figure}[!t]
\centering
\includegraphics[width=0.5\textwidth]{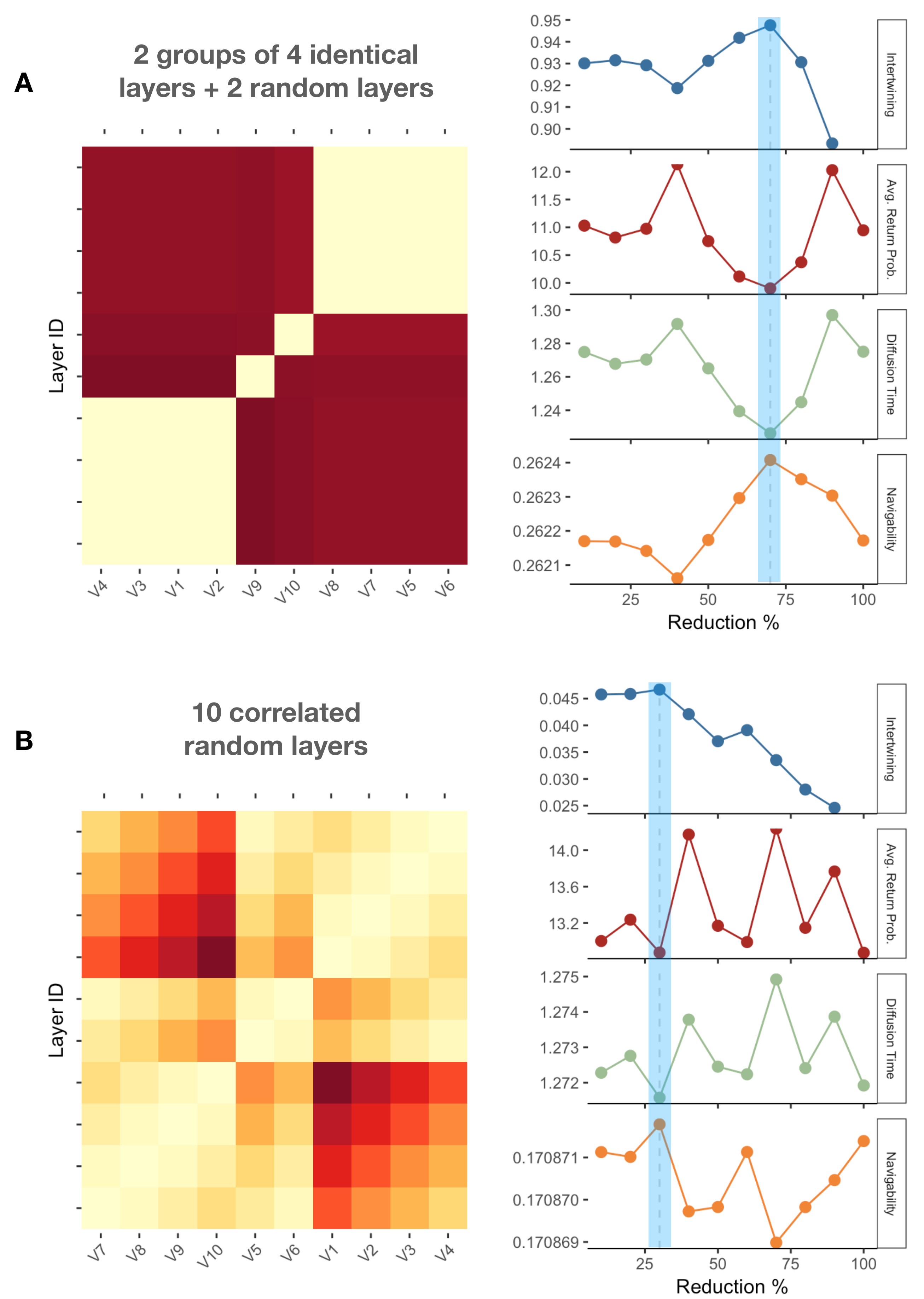}
\caption{\label{fig:app:synthetic}\textbf{Functional reduction of additional synthetic systems.} See Appendix~\ref{app:syntheticapp} for details about the two benchmarks.}
\end{figure}

\section{Relating intertwining to transport properties}\label{app:intertwining_transport}

Let us define $\lambda_{2}^{(\ell)} = \lambda_{2} + \Delta \lambda_{2}^{(\ell)}$, from which we have in the limit of large $\tau$:
\begin{eqnarray}
\zint &\approx& 1 + e^{-\tau \lambda_{2}} \nonumber\\
Z^{(\ell)} &\approx& 1 + e^{-\tau \lambda_{2}^{(\ell)}} = 1 + e^{-\tau (\lambda_{2} + \Delta\lambda_{2}^{(\ell)}) } \approx \zint + \Delta Z^{(\ell) },\nonumber 
\end{eqnarray}
where 
\begin{eqnarray}
\Delta Z^{(\ell) } = -\tau e^{-\tau \lambda_{2}} \Delta\lambda_{2}^{(\ell)}.
\end{eqnarray}
We have already showed that for large $\tau$ the fundamental inequality reduces to
\begin{eqnarray}
\zint &\leq& \zint + \frac{1}{L} \sum_{\ell=1}^{L} \Delta Z^{(\ell)},
\end{eqnarray}
leading to
\begin{eqnarray}
\zint \leq \zint -\tau e^{-\tau \lambda_{2}} \frac{1}{L} \sum\limits_{\ell=1}^{L}\Delta\lambda_{2}^{(\ell)},
\end{eqnarray}
or equivalently:
\begin{eqnarray}
0 \leq -\tau e^{-\tau \lambda_{2}} \overline{\Delta\lambda_{2}^{(\ell)}},
\end{eqnarray}
from which we deduce that $\overline{\Delta\lambda_{2}^{(\ell)}} \leq 0 \leq \lambda_2$. The resulting intertwining becomes
\begin{eqnarray}
\mathcal{I}(\tau) =(\tau+1) \frac{\overline{\Delta Z^{(\ell)}}(\tau)}{\zint(\tau)}=(\tau+1)\frac{ -\tau e^{-\tau \lambda_{2}} \overline{\Delta\lambda_{2}^{(\ell)}} }{\zint} \nonumber
\end{eqnarray}
and for the relative intertwining we have
\begin{eqnarray}
\mathcal{I}^{\star}(\tau)&=&\frac{\frac{ -\tau e^{-\tau \lambda_{2}} \overline{\Delta\lambda_{2}^{(\ell)}} }{\zint}}{\frac{\zint-1}{\zint}+\frac{ -\tau e^{-\tau \lambda_{2}} \overline{\Delta\lambda_{2}^{(\ell)}} }{\zint}}\nonumber \\
&=& \frac{-\tau \overline{\Delta\lambda_{2}^{(\ell)}}}{1 -\tau \overline{\Delta\lambda_{2}^{(\ell)}}} \nonumber \\
&\approx& -\tau \overline{\Delta\lambda_{2}^{(\ell)}} = - \tau (  \overline{\lambda_{2}^{(\ell)}} - \lambda_{2} ) = \tau(\lambda_{2} - \overline{\lambda_{2}^{(\ell)}} ).\nonumber
\end{eqnarray}
The last step highlights the existence of a direct relationship between our quality function for quantifying functional redundancy and the second smallest eigenvalue of the Laplacian matrix, $\lambda_{2}$. It is easy to show that -- regardless of the value of $\tau$ -- the aggregation step $m=m_{opt}$ where $\mathcal{I}^{\star}(m;\tau)$ is globally maximum corresponds to a minimum in the characteristic diffusion time, quantified by $1/\lambda_2$ and the navigability, which at first-order is proportional to $-\lambda_2$.

\section{Additional synthetic benchmarks}\label{app:syntheticapp}

In Fig.~\ref{fig:app:synthetic}A we consider a multiplex network with 100 nodes, whose layers are all random networks with wiring probability $p=0.2$, with the additional constraint that two bunch of layers are exactly the same. Figure~\ref{fig:app:synthetic}B shows results for a multiplex network of 100 nodes and 10 layers, in which the first layer is a random networks with $p=0.4$, the second layer is a copy of the first but after randomly cutting links with a probability of $4/100$, and so forth so on, to induce correlations among random layers. The functional analysis of these systems is in agreement with our expectation: layers which are exactly the same are coupled together: the resulting groups are noisy and independent networks, which can not be further aggregated.

\clearpage

\bibliographystyle{apsrev4-1}
\bibliography{biblio}

\end{document}